%% file: Main.tex
\documentclass{jfm}

\usepackage[]{amsmath}
\usepackage[]{xcolor}
\usepackage[]{subfigure}
\usepackage{upgreek}
\usepackage{overpic}
\usepackage[]{multirow}
\usepackage[]{url}
\usepackage[]{nicefrac}
\usepackage{bm}

\graphicspath{{imgs/}}

\DeclareMathOperator*{\argmin}{argmin}

\begin{document}

\newtheorem{lemma}{Lemma}
\newtheorem{corollary}{Corollary}

\shorttitle{Sparse reduced-order modeling} 
\shortauthor{J.-Ch.~Loiseau, B.~R.~Noack and S.~L.~Brunton} 

\title{Sparse reduced-order modeling: Sensor-based dynamics to full-state estimation}

\author{Jean-Christophe Loiseau\aff{1} \corresp{\email{loiseau.jc@gmail.com}},
	Bernd R.~Noack\aff{2,3,4}
        \and Steven L.~Brunton\aff{5}}

\affiliation
{
\aff{1}
Laboratoire DynFluid, Arts et M\'etiers ParisTech, 75013 Paris, France
\aff{2}
Laboratoire d'Informatique pour la M\'ecanique et les Sciences de l'Ing\'enieur,
LIMSI-CNRS,  Rue John von Neumann, Campus Universitaire d'Orsay, B\^at 508,
F-91403 Orsay, France
\aff{3}
Institut f\"ur Str\"omungsmechanik,
Technische Universit\"at Braunschweig,
Hermann-Blenk-Stra{\ss}e 37,
D-38108 Braunschweig, Germany
\aff{4}
Institut f\"ur Str\"omungsmechanik und Technische Akustik (ISTA),
Technische Universit\"at Berlin,
M\"uller-Breslau-Stra{\ss}e 8,
D-10623 Berlin, Germany
\aff{5}
Department of Mechanical Engineering, University of Washington, Seattle, WA 98195, USA
}

\maketitle

\input{S0_Abstract.tex}

\input{S1_Introduction.tex}

\input{S2_Configuration.tex}

\input{S3_Modeling.tex}

\input{S4_Results.tex}

\input{S5_Conclusions.tex}

\appendix
\input{SA_Appendices.tex}

\bibliography{bibliography,Main_Bernd}
\bibliographystyle{jfm}

\end{document}

%% file: S0_Abstract.tex
\begin{abstract}
We propose a general dynamic reduced-order modeling framework
for typical experimental data: 
time-resolved sensor data and optional non-time-resolved PIV snapshots.
This framework contains four steps.
First, the sensor signals are lifted to a dynamic feature space.
Second, we identify a sparse human-interpretable 
nonlinear dynamical system for the feature state based on the sparse identification of nonlinear dynamics (SINDy).
Third, if PIV snapshots are available, a local linear mapping from the feature state to velocity fields 
is shown to be orders of magnitudes more accurate 
than optimal modal expansions of the same order.  
Fourth, a generalized feature-based modal decomposition identifies coherent structures that are most dynamically correlated with the linear and nonlinear interaction terms in the sparse model, adding interpretability.  
Steps 1 and 2 define a black-box model.
Optional steps 3 and 4 lift the black-box dynamics 
to a `gray-box' model of the coherent structures, if non-time-resolved full-state data is available. 
This gray-box modeling strategy is successfully applied 
to the transient and post-transient laminar cylinder wake,
and compares favorably with a POD model. 
We foresee numerous applications of this highly flexible modeling strategy,
including estimation, prediction and control. 
Moreover, the feature space may be based on intrinsic coordinates, 
which are unaffected by a key challenge of modal expansion: 
the slow change of low-dimensional coherent structures with changing geometry and varying parameters.
\end{abstract}

%% file: S1_Introduction.tex
\section{Introduction}
\label{ToC:Intro}

Understanding and modeling complex fluid flows is a central focus in many scientific, technological, and industrial applications, including energy (e.g., wind, tidal, and combustion), transportation (e.g., planes, trains, and automobiles), security (e.g. airborne contamination), and medicine (e.g., artificial hearts and artificial respiration).
Improved models of engineering flows have the potential to dramatically improve performance in these systems through optimization and control, resulting in practical gains such as drag reduction, lift increase, and mixing enhancement~\citep{fabbiane2014amr,Brunton2015amr,Sipp2016amr,arfm:rowley:2016}.
Although the Navier-Stokes equations provide a detailed mathematical model, this representation may be difficult to use for engineering design, optimization, and control.
\cite{Feynman2013book} points out the limitation of the governing equations to reveal underlying behavior:
\begin{quotation}
``The test of science is its ability to predict.  Had you never visited the earth, could you predict the thunderstorms, the volcanos, the ocean waves, the auroras, and the colorful sunset?''
\end{quotation}

Instead of studying the Navier-Stokes equations directly, fluid systems are commonly discretized into a high-dimensional, nonlinear dynamical system with many degrees of freedom and multi-scale interactions.
These equations are expensive to simulate, making them unwieldy for iterative optimization or in-time control, and they may also obscure the underlying physics, which often evolves on a low-dimensional attractor~\citep{HLBR_turb,jfm:noack:2003}. The various fidelities of model description were described by \citet{Wiener1948book}:  `white-box' describes an accurate evolution equation based on first principles (e.g., Navier-Stokes discretization), `gray-box' describes a low-dimensional model approximating the full state (e.g., POD-Galerkin models), and `black-box' describes input--output models that lack a connection to the full state space (e.g., neural networks).

In the following, we outline related reduced-order models as our point of departure in \textsection \ref{ToC:Intro:ROM} and foreshadow proposed innovations of this study in \textsection \ref{ToC:Intro:GBM}.

\subsection{Related reduced-order models as point of departure}
\label{ToC:Intro:ROM}

Reduced-order models provide minimal descriptions of the underlying fluid behavior in a compact and computationally efficient representation.
There are many techniques for reduced-order modeling, ranging from physical reductions to purely data-driven methods, and nearly everything in between.
Proper orthogonal decomposition (POD)~\citep{qam:sirovich:1987,berkooz1993proper,HLBR_turb} provides a low-rank modal decomposition of fluid flow field data, extracting the most energetic modes.
It is then possible to Galerkin project the Navier-Stokes equations onto these modes, resulting in an approximate, low-dimensional model in terms of mode coefficients~\citep{Noack2011book,Carlberg2015siamjsc}.
POD-Galerkin models are widely used, as they are interpretable, gray-box models, and it is straightforward to reconstruct the high-dimensional flow field from the low-dimensional model via POD modes. The first pioneering example of \citet{Aubry1988jfm} features wall turbulence---almost three decades ago.
Subsequent POD models have been developed
for the transitional boundary layer \citep{Rempfer1994jfm2},
the mixing layer \citep{Ukeiley2001jfm,Wei2009jfm},
the cylinder wake \citep{Deane1991pfa,Galletti2004jfm}, and
the Ahmed body wake \citep{Osth2014jfm}, to name only a few.

POD-Galerkin modeling is challenging for
 changing domains \citep{Bourguet2011jcp},
changing boundary conditions \citep{Graham1999ijnme}
and slow deformation of the modal basis \citep{Babaee2016ptrs}.
Standard Galerkin projection can also be expected to suffer from stability issues~\citep{Rempfer2000tcfd,Schlegel2015jfm,Carlberg2017jcp},
although including energy-preserving constraints may improve
the long-time stability and performance of nonlinear models \citep{Balajewicz2013jfm,Cordier2013ef}.
POD-Galerkin models  tend to be  valid for a narrow range of operating conditions, namely around the data set used for the POD modes.
Transients also pose a challenge to POD modeling.
\cite{jfm:noack:2003} and~\cite{pof:tadmor:2010} demonstrate the ability of a low-dimensional model to reproduce nonlinear transients of the von K\`arm\`an vortex shedding past a two-dimensional cylinder, provided the projection basis includes a \emph{shift mode} quantifying the distortion between the linearly unstable base flow and marginally stable mean flow.
These techniques have been extended
to include the effect of wall actuation \citep{Graham1999ijnme,Rediniotis2002jfe}.

In addition to the physics-based Galerkin projection, data-driven modeling approaches are prevalent in fluid dynamics~\citep{Brunton2015amr,arfm:rowley:2016}.
For example, dynamic mode decomposition (DMD)~\citep{jfm:schmid:2010,Rowley2009jfm,Kutz2016book}, the eigensystem realization algorithm (ERA)~\citep{jgcd:juang:1985}, Koopman analysis~\citep{Mezic2005nd,Mezic2013arfm,Tu2014jcd,Williams2015jnls}, cluster-based reduced order models (CROM)~\citep{Kaiser2014jfm}, NARMAX models~\citep{Billings2013book,Semeraro2016arxiv,Zhang2012aiaa,Glaz2010aiaa}, and network analysis~\citep{nair2015network} have all been used to identify dynamical systems models from fluids data, without relying on knowledge of the underlying Navier-Stokes equations.
DMD models are readily obtained directly from data, and they provide interpretability in terms of flow structures, but the resulting models are linear, and the connection to nonlinear systems is tenuous unless DMD is enriched with nonlinear functions of the data~\citep{Williams2015jnls,Kutz2016book}.
Neural networks have long been used for flow modeling and control~\citep{Milano2002jcp,Zhang2015aiaa,Lee1997pof,Krizhevsky2012nips}, and recently deep neural networks have been used for Reynolds averaged turbulence modeling~\citep{Ling2016jfm,Kutz2017jfm}.
Parsimony has also become an overarching theme when using machine learning to model nonlinear dynamics.
\cite{Bongard2007pnas} and \cite{Schmidt2009science} discover governing dynamics and conservation laws using genetic programming along with a Pareto analysis to balance model accuracy and complexity.
However, some machine learning methods, such as neural networks and genetic programming, may be prone to overfitting, have limited interpretability, and make it difficult to incorporate known physical constraints.

Recently, \cite{pnas:brunton:2016} introduced the sparse identification of nonlinear dynamics (SINDy), which identifies parsimonious nonlinear models from data.
SINDy follows the principle of Ockham's razor, resting on the assumption that there are only a few important terms that govern the dynamics of a system, so that the equations are sparse in the space of possible functions.
Sparse regression is then used to efficiently determine the fewest terms in the dynamics required to accurately represent the data, preventing overfitting.
Because SINDy is based on linear algebra (i.e., the nonlinear dynamics are represented as a linear combination of candidate nonlinear functions), the method is readily extended to incorporate known physical constraints~\citep{Loiseau2016arxiv}.
In general, it is possible to obtain nonlinear models using genetic programming or SINDy on POD or DMD mode coefficients, which make these methods \emph{gray} box, having a transformation from the model back to the high-dimensional, interpretable state-space.
However, models developed on POD/DMD mode coefficients
may still suffer from fundamental challenges
of traditional POD-Galerkin models,
such as capturing changing boundary condition,
moving geometry, and varying operating condition.

\subsection{Contribution of this work}
\label{ToC:Intro:GBM}

In this work, we introduce a new gray-box modeling procedure
that yields interpretable nonlinear models from measurement data.
The method is applied to the well-investigated two-dimensional transient flow past a circular cylinder
with slow change of the base flow
and varying coherent structures \citep{Tadmor2011ptrsa}.
In particular, we develop nonlinear models only from lift measurements
that accurately capture steady-state and transient flow behavior.
First, a feature vector is constructed from the lift signal,
including a time-delayed value.
Second, a sparse dynamical model is identified in this feature space.
For the following steps, full-state measurement data is assumed to be available.
Third, a local linear mapping from feature vector to velocity field is constructed
using a K-nearest neighbors (KNN) approach.
This mapping provides significantly more accurate flow reconstruction,
as compared to a POD-Galerkin model of the same order.
Technically,  we mitigate the significant challenges
of using an `elliptic' Galerkin modeling approach to an `hyperbolic' dynamics:
The global modes  connect all locations in all directions instantaneously
while the coherent structures are convected with the flow \citep{Noack2016jfm2}.

Furthermore,
it is also possible to construct a generalized set of modes
that are most dynamically correlated with the given terms in the identified model.
The resulting modal decomposition is similar in concept to DMD,
although the present approach corrects linear modes
for the effect of nonlinearity and also reveals
entirely new structures associated with given nonlinear interaction terms.
With the associated full-state data and generalized modes,
it is possible to reconstruct the full-state associated
with a given low-dimensional prediction in the reduced-order gray-box model.

\begin{figure}
\begin{center}
\vspace{.1in}
\begin{overpic}[width=\textwidth]{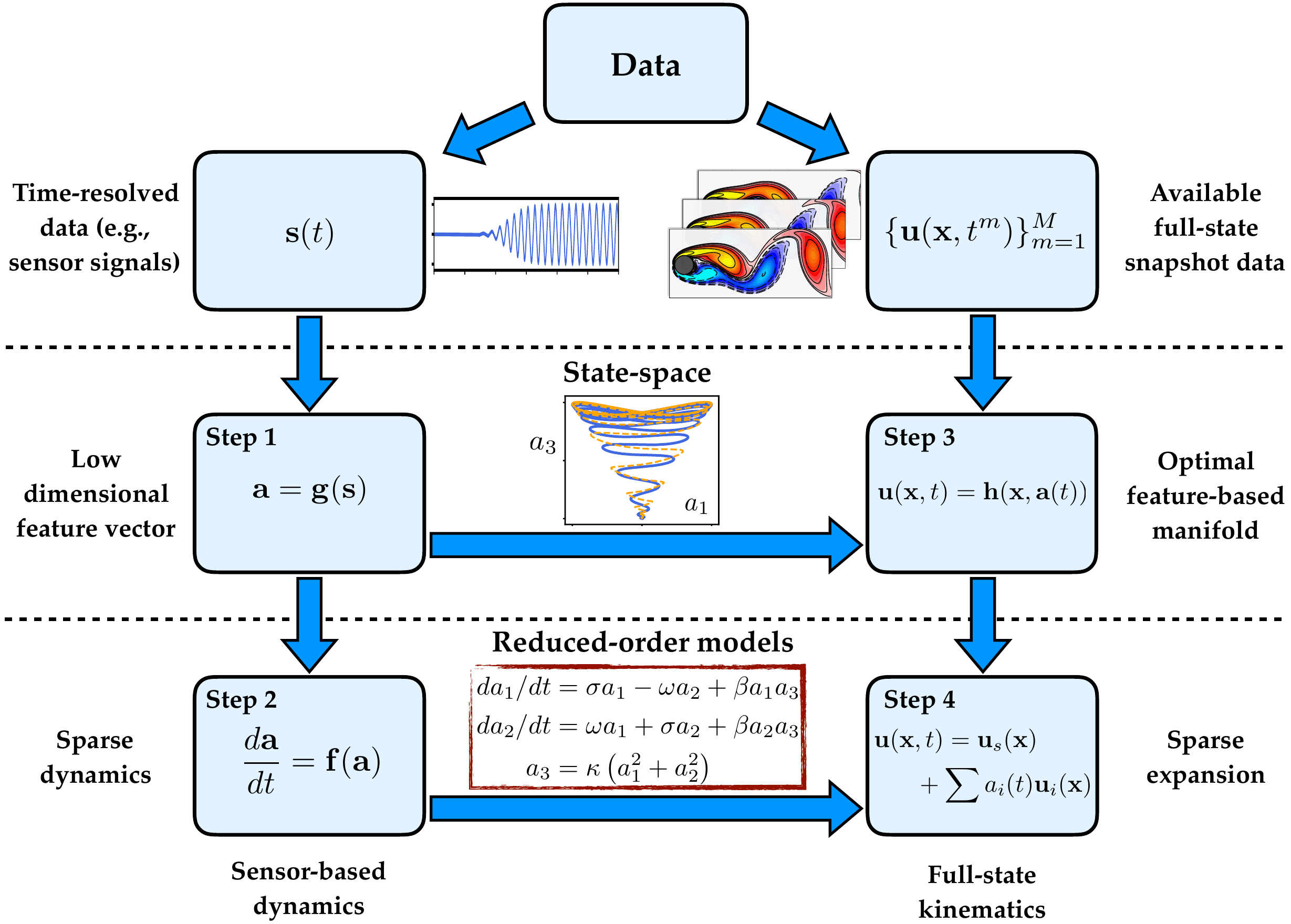}
\end{overpic}
\vspace{-.2in}
\caption{Schematic overview of the proposed sparse modeling procedure.  A sparse dynamical system~\citep{pnas:brunton:2016} is identified based on features obtained from sensor signals $\mathbf{s}$, and the full state $\mathbf{u}$ may also be estimated with the availability of PIV snapshots (optional). }\label{Fig:overview}
\vspace{-.2in}
\end{center}
\end{figure}

To summarize, the resulting gray-box modeling procedure
has the following beneficial features:  (i) it captures nonlinear physics, (ii) it is based on a simple, non-invasive computational algorithm, (iii) the resulting model is interpretable in terms of nonlinear interaction physics and generalized modes (optional with full-state data), and (iv) modeling feature vectors is more robust to mode deformation, moving geometry, and varying operating condition. This procedure is shown schematically in figure \ref{Fig:overview}.

The manuscript is organized as follows: \textsection \ref{section:experiment} provides an overview of the flow configuration considered in this work, namely the incompressible, two-dimensional flow past a circular cylinder at $\Rey=100$.  \textsection \ref{section:graybox} describes the proposed gray-box modeling procedure, including modeling in feature space and obtaining a generalized modal expansion if full-state data is available.  In \textsection \ref{section:results}, numerical results for the gray-box modeling procedure are presented and analyzed for the cylinder flow.  Finally, \textsection \ref{section:conclusions} summarizes our key findings and provides the reader with possible future directions to extend this work.

%
%

%% file: S2_Configuration.tex
\section{Flow configuration}\label{section:experiment}

The flow configuration considered in the present work is the two-dimensional incompressible viscous flow past a circular cylinder at ${\Rey = 100}$. This Reynolds number, based on the free-stream velocity $U_{\infty}$, the cylinder diameter $D$ and the kinematic viscosity $\nu$, is well above the onset of vortex shedding~\citep{jem:zebib:1987, jfm:schumm:1994} and below the onset of three-dimensional instabilities~\citep{pof:zhang:1995, jfm:barkley:1996}. In the fluid dynamics community, a large body of literature exists in which this particular setup has been chosen to illustrate modal decomposition~\citep{jfm:bagheri:2013} and model identification techniques~\citep{jfm:noack:2003, pre:sengupta:2015, pnas:brunton:2016, arfm:rowley:2016}. This setup is thus a particularly compelling test case to illustrate our model identification strategy, as well as to draw connections and quantify its performance against other well-established techniques.

The dynamics of the flow are governed by the incompressible Navier-Stokes equations
\begin{equation}
	\begin{aligned}
		& \frac{\partial {\bm u}}{\partial t} + ({\bm u} \cdot \nabla){\bm u} = -\nabla p + \frac{1}{\Rey} \nabla^2 {\bm u} \\
		& \nabla \cdot {\bm u} = 0,
	\end{aligned}
	\label{eq: Navier-Stokes}
\end{equation}
where ${\bm u} = (u, v)^T$ and $p$ are the velocity and pressure fields, respectively. The center of the cylinder has been chosen as the origin of the reference frame ${\bm x} = (x, y)$, where $x$ denotes the streamwise coordinate and $y$ denotes the spanwise coordinate.
This study considers the same computational domain as in~\cite{jfm:noack:2003}, extending from $x=-5$ to $x=15$ in the streamwise direction, and from $y=-5$ to $y=5$ in the spanwise direction.
A uniform velocity profile is prescribed at the inflow, a classical stress-free boundary condition is used at the outflow, and free-slip boundary conditions are used on the lateral boundaries of the computational domain.
Based on the spectral element solver Nek 5000 \citep{nek5000_site}, the domain is discretized by 1832 seventh-order spectral elements.
Finally, the time-integration of the diffusive terms relies on a Backward Differentiation of order 3 (BDF3), while the convective terms are advanced in time based on a third-order accurate extrapolation.

\begin{figure}
\includegraphics[scale=1]{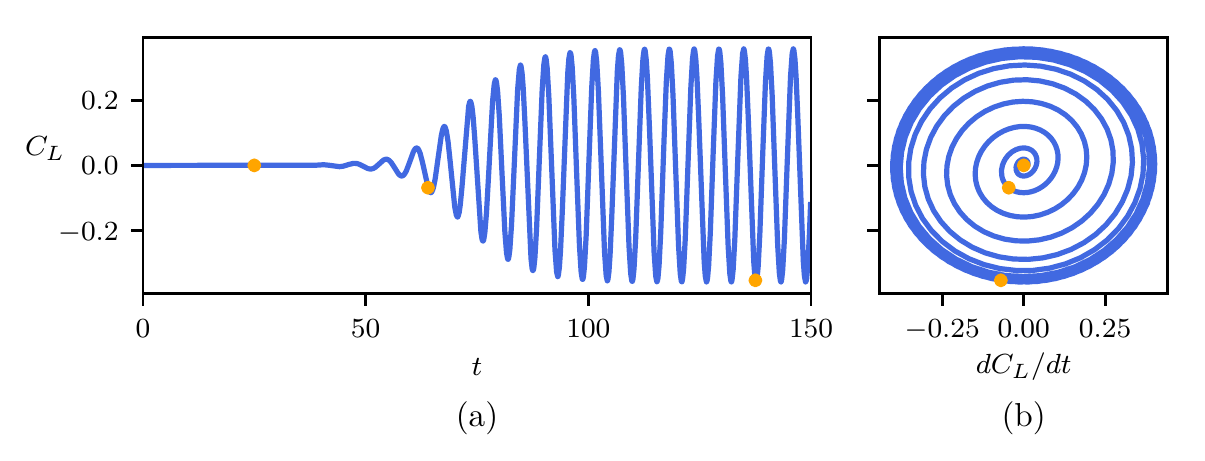}
\caption{(a) Evolution of the lift coefficient $C_L$ as a function of time for the two-dimensional cylinder flow at $Re=100$. (b) Trajectory of the system in the phase-plane ($C_L$, $d C_L/dt$). In both figures, the orange dots indicate the instants of time for which the corresponding vorticity field is shown in figure~\ref{fig: flow snapshots}.}
\label{fig: lift signal}
\end{figure}

\begin{figure}
\includegraphics[width=\textwidth]{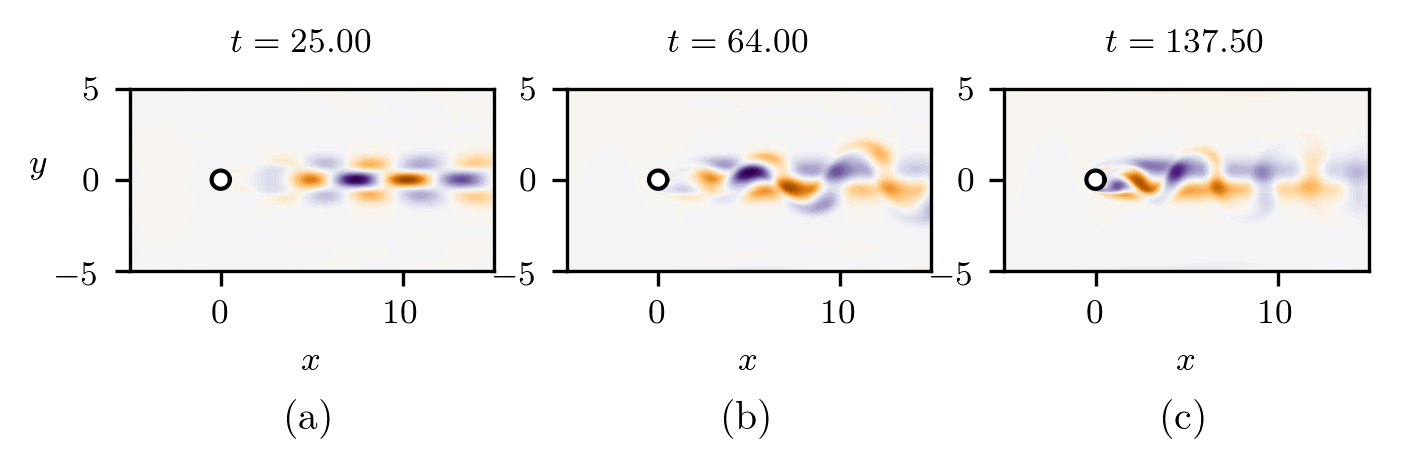}
\caption{Snapshots of the vorticity field at the different time instants highlighted in figure~\ref{fig: lift signal}. Note that the vorticity field of the linearly unstable baseflow, ${\bm u}_s({\bm x})$, has been subtracted in order to highlight the vorticity induced by the perturbation.}
\label{fig: flow snapshots}
\end{figure}

Many of the direct numerical simulations performed in this work have been initialized with the following initial condition
\begin{equation}
	{\bm u}({\bm x}, 0) = {\bm u}_s({\bm x}) + 0.001 {\bm \epsilon}({\bm x}),
\end{equation}
where ${\bm u}_s({\bm x})$ is the linearly unstable steady solution of the Navier-Stokes equations and ${\bm \epsilon}({\bm x})$ is a zero-mean and unit-variance random white-noise velocity field.
Each simulation is run for 150 convective time units, providing $M = 1200$ equidistantly sampled velocity snapshots ${\bm u}^m({\bm x}) = {\bm u}({\bm x}, t^m)$, $m = 1, \cdots, M$, and associated measurements of the lift and drag coefficients, $C_L(t^m)$ and $C_D(t^m)$.
This time-span covers the entire unforced transient phase, from the steady solution to the fully developed von K\'arm\'an vortex street.
Figure \ref{fig: lift signal} depicts a typical evolution of the lift coefficient $C_L$, while figure \ref{fig: flow snapshots} shows snapshots of the vorticity field at different instants of time.
For $t \le 50$, the flow is governed by linear dynamics.
Consequently, the vorticity field of the perturbation ${\bm v}({\bm x}, t) = {\bm u}({\bm x}, t) - {\bm u}_s({\bm x})$, as shown in figure \ref{fig: flow snapshots}(a), can be well approximated by the leading instability mode.
For $50 \le t \le 80$, the perturbation grows to an extent that nonlinear effects cause the perturbation to distort.
Eventually, for $t \ge 80$, the flow settles onto a periodic limit cycle corresponding to the onset of the von K\`arm\`an vortex street.
Given the evolution of the lift coefficient depicted in figure \ref{fig: lift signal} and the associated snapshots shown in figure \ref{fig: flow snapshots}, the aim of the present work is to propose a new reduced-order modeling strategy able to accurately reproduce such dynamics and flow structures, but without the need for the full flow field or modes.

%% file: S3_Modeling.tex
\section{Sparse sensor-based modeling}\label{section:graybox}

Here, we discuss the core mathematical and algorithmic framework used to identify nonlinear reduced-order models and build a full-state estimator from data. The identification procedure relies on the \emph{Sparse Identification of Nonlinear Dynamics} (SINDy) method \citep{pnas:brunton:2016} briefly summarized in \textsection \ref{subsec: SINDy}. As a second step, we introduce the reader to two different full-state estimators in \textsection \ref{subsec: full-state estimation}.

\subsection{From sensor signals to feature space}
\label{subsec: From sensor space to feature space}

Experimentalists typically only have access to a limited number
of time-resolved sensor measurements ${\bm s}$.
In contrast,
one can gather space-time resolved data at every single point within the computational domain when performing direct numerical simulations.
The aim of the present work is to illustrate
how one can leverage recent advances in system identification
and machine learning to construct reduced-order models directly from limited sensor measurements.
We mimic these experimental conditions with direct numerical simulations.
In the present work, we consider a sensor measurement given by
\begin{equation}
\begin{aligned}
s & := C_L(t),
\end{aligned}
\end{equation}
where $C_L$ is the lift coefficient.
Note that in general ${\bm s}$ may be a vector of measurements, for example including the lift and drag coefficients, pressure measurements on an immersed body, or point velocity field measurements at select locations; however, in the present study, we will show that lift is sufficient to characterize the flow.
Given the sensor measurements ${\bm s}$, our aim is to identify a low-order model that allows us to predict the evolution of our system.
However, the sensor measurements may need to be augmented, or \emph{lifted}, to include functions of the sensor measurements.
We consider the augmented state ${\bm a}$ to be a feature vector given by
\begin{equation}
\bm{a} = {\bm g}({\bm s}).
\end{equation}
There are many choices for the mapping ${\bm g}$ to enrich the sensor measurements and improve models.
If the sensors are sufficient to define the state of the system, then ${\bm g}$ may be the identity map.
If the sensors consist of high-dimensional snapshots, then ${\bm g}$ may extract POD mode coefficients.
It may also be possible to augment the measurements with delay-embedding~\citep{takens:1981,jgcd:juang:1985,Brunton2016havok}.
More generally, choosing a good transformation, ${\bm g}$, is an important open problem, with connections to representation theory and the Koopman operator perspective on dynamical systems~\citep{Mezic2005nd,Mezic2013arfm,Williams2015jnls,Brunton2016plosone,Brunton2016havok,Arbabi2016arxiv}.
In the present study, we choose ${\bm g}$ to augment the sensor measurement of the lift coefficient with its time derivative, along with a proper scaling:
\begin{equation}
\begin{aligned}
a_1 & := s(t), \\
a_2 & := \frac{1}{\omega_{\infty}} \frac{ds}{dt}(t),
\end{aligned}
\label{eq: sensor measurement}
\end{equation}
where $\omega_{\infty}$ is the post-transient angular shedding frequency.
This choice has been guided by physical considerations:
the lift can be measured and allows for the characterization of the state of the flow.
It is also well-known that the flow around a two-dimensional cylinder at $\Rey=100$ only necessitates three degrees of freedom~\citep{jfm:noack:2003}, or features, to be approximately described:
the shedding amplitude, shedding phase and the degree of base flow deformation.
In this work, we will show that the degree of base flow deformation can be described by the drag coefficient, which is then modeled as an algebraic equation of the feature vector ${\bm a}$.

This setup mimics an experiment in \cite{Hosseini2016jfm}.
The pressure difference between the top and bottom side of a cylinder
is a surrogate quantity for the lift.
This difference  has been recorded in a time-resolved manner
while non-time-resolved PIV flow snapshots were taken.
The pressure difference and a flow estimator
have been employed for a time-resolved estimate of the fluid flow field,
comprising the base flow,
von K\'arm\'an vortex shedding and the second harmonics.
The analogues of $a_1$ and $a_2$ were the cosine and sine Morlet transforms of the pressure difference history.

\subsection{Sparse Identification of Nonlinear Dynamics}
\label{subsec: SINDy}

Identifying reduced-order models from data is a central challenge in mathematical physics, with a rich history of developments in fluid dynamics.
The form of the dynamics is typically either constrained via prior knowledge, as in the Galerkin projection, or a particular model structure is chosen heuristically, and parameters are optimized to match the data.
Simultaneous identification of the model structure and parameters from data is considerably more challenging, as there are combinatorially many possible model structures.
The sparse identification of nonlinear dynamics (SINDy) architecture~\citep{pnas:brunton:2016} bypasses the intractable combinatorial search through all possible model structures, leveraging the fact that many systems may be modeled by dynamics ${\bm f}$ that are sparse in the space of possible right-hand side functions:
\begin{equation}
\displaystyle \frac{\mathrm{d} \bm{a}}{\mathrm{d} t}  = {\bm f}(\bm{a}),
\label{Eq:Dynamics}
\end{equation}
where $\bm{a}$ is the same state vector as in \textsection \ref{subsec: From sensor space to feature space}.
It is then possible to solve for the relevant terms
that are active in the dynamics using either a convex $\ell_1$-regularized regression~\citep{Tibshirani1996lasso} or a sequentially thresholded least-squares~\citep{pnas:brunton:2016}; these algorithms penalize the number of terms in the dynamics and scale favorably to large problems.

First, time-series data is collected and formed into a data matrix:
\begin{equation}
{\bf A} = \begin{bmatrix} \bm{a}(t^1) & \bm{a}(t^2) & \cdots & \bm{a}(t^M)\end{bmatrix}^T
\end{equation}
where `$T$' denotes the matrix transpose.
A similar matrix of derivatives is formed:
\begin{equation}
\dot{{\bf A}} = \begin{bmatrix} \displaystyle \frac{\mathrm{d}\bm{a}}{\mathrm{d}t}(t^1) & \displaystyle \frac{\mathrm{d}\bm{a}}{\mathrm{d}t}(t^2) & \cdots & \displaystyle \frac{\mathrm{d}\bm{a}}{\mathrm{d}t}(t^M)\end{bmatrix}^T.
\label{Eq:Xdot}
\end{equation}
In practice, this may be computed directly from the data in ${\bf A}$.
However, for noisy data, the total-variation regularized derivative \citep{Chartrand2011isrnam} tends to provide numerically robust derivatives.
Based on the data in ${\bf A}$, a library of candidate nonlinear functions $\boldsymbol{\Uptheta}({\bf A})$ is constructed:
\begin{eqnarray}
\boldsymbol{\Uptheta}({\bf A}) = \begin{bmatrix} \mathbf{1} & {\bf A} & {\bf A}^2 & \cdots & {\bf A}^d  & \cdots &   \sin({\bf A}) & \cdots  \end{bmatrix}.
\label{Eq:NLLibrary}
\end{eqnarray}
Here, the matrix ${\bf A}^d$ denotes a matrix with column vectors given by all possible time-series of $d$-th degree polynomials in the state $\bm{a}$.
The dynamical system in Eq.~\eqref{Eq:Dynamics} may now be represented in terms of the data matrices in Eqs.~\eqref{Eq:Xdot} and~\eqref{Eq:NLLibrary} as
\begin{eqnarray}
\dot{{\bf A}} = \boldsymbol{\Uptheta}({\bf A})\boldsymbol{\Upxi}.\label{Eq:SINDy1}
\end{eqnarray}
Each column $\boldsymbol{\Upxi}_k$ in $\boldsymbol{\Upxi}$ is a vector of coefficients determining the active  terms in the $k$-th row equation in Eq.~\eqref{Eq:Dynamics}.
A parsimonious model will provide an accurate model fit in Eq.~\eqref{Eq:SINDy1} with as few terms as possible in $\boldsymbol{\Upxi}$.
Such a model may be identified using a convex $\ell_1$-regularized sparse regression:
\begin{eqnarray}
\boldsymbol{\Upxi}_k = \argmin_{\boldsymbol{\Upxi}_k'}\|\dot{\mathbf{A}}_k - \boldsymbol{\Uptheta}(\mathbf{A})\boldsymbol{\Upxi}_k'\|_2+\lambda \|\boldsymbol{\Upxi}_k'\|_1.\label{Eq:SparseRegression}
\end{eqnarray}
Here, $\dot{\mathbf{A}}_k$ is the $k$-th column of $\dot{\mathbf{A}}$.
Sparse regression, such as the LASSO~\citep{Tibshirani1996lasso} or the sequential thresholded least-squares algorithm used in SINDy, improves the numerical robustness of this identification for noisy overdetermined problems, in contrast to earlier methods~\citep{Wang2011prl} that used compressed sensing~\citep{Donoho2006ieeetit,Candes2006picm}.
Once identified, the sparse vectors $\boldsymbol{\Upxi}_k$ may be synthesized into a nonlinear dynamical system model:
\begin{eqnarray}
\frac{\mathrm{d} a_k}{\mathrm{d} t} = \boldsymbol{\Uptheta}(\bm{a})\boldsymbol{\Upxi}_k,
\end{eqnarray}
where  $a_k$ is the $k$-th element of $\bm{a}$ and $\boldsymbol{\Uptheta}(\bm{a})$ is a row vector of symbolic functions of $\bm{a}$, as opposed to the data matrix $\boldsymbol{\Uptheta}(\mathbf{A})$.
Identifying the most parsimonious nonlinear model by applying sparse regression in the library $\boldsymbol{\Uptheta}$ is a convex procedure.
The alternative approach, which involves regression onto every possible sparse nonlinear structure, constitutes an intractable brute-force procedure.
SINDy thus bypasses this combinatorial search with modern convex optimization and machine learning.
The SINDy algorithm is closely related to NARMAX models~\citep{Billings2013book} and fast function extraction (FFX)~\citep{Mcconaghy2011book}.

A major benefit of the SINDy architecture is its ability to identify parsimonious models that contain only the required nonlinear terms, resulting in interpretable models that avoid overfitting.
In the optimization above, the sparsifying parameter $\lambda$ may be varied from $\lambda=0$ (i.e., least-squares) to $\lambda\rightarrow\infty$ (i.e, trivial dynamics $\mathrm{d}\bm{a}/\mathrm{d}t = \mathbf{0}$), sweeping out a Pareto front.
To identify the most parsimonious model that best balances model complexity with accuracy, \cite{arxiv:mangan:2016} proposed an efficient methodology to rank candidate models on the Pareto front using the Akaike information criterion (AIC)~\citep{Akaike1974ieeetac} or the Bayes information criterion (BIC)~\citep{Schwarz1978BIC}.

The embedding of nonlinear dynamics in terms of a linear regression problem in \eqref{Eq:SparseRegression} makes the SINDy method highly extensible.
Recent extensions to SINDy enable the identification of nonlinear differential equations with rational function nonlinearities by reformulating the problem as an implicit differential equation and solving for the active terms by finding the sparsest vector in the null space of an augmented library containing functions of the state and derivative terms~\citep{Mangan2016ieee}.
SINDy has also been generalized to identify partial differential equations from data~\citep{Rudy2016arxiv,Schaeffer2017prsa}, and has been extended to include inputs and control~\citep{Brunton2016nolcos}.

\subsection{Full-state estimation}
\label{subsec: full-state estimation}

The proposed methodology enables the identification of a low-order model that reproduces the system dynamics recorded by a few sensors.
Although it may provide useful insights into the physics, it does not allow for a straightforward full-state estimation of the system considered.
To accommodate this estimation, one needs to define a function
\begin{equation}
\bm{h}(\bm{x}, \bm{a}) \approx \bm{u}(\bm{x}, t)
\label{eq: full-state estimation}
\end{equation}
mapping the state of the system from the low-dimensional feature space to the high-dimensional physical space.
In the most general case, $\bm{h}$ is a nonlinear mapping function.
Note however that $\bm{h}(\bm{x},\bm{0})$ physically corresponds to the reference state $\bm{u}_s(\bm{x})$ at $\bm{a}=0$.
While this reference state is classically chosen as the mean flow, in the present work, it is chosen as the linearly unstable steady solution $\bm{u}_s(\mathbf{x})$ of the Navier-Stokes equations.
In the following, two different strategies to approximate $\bm{h}({\bm x}, \bm{a})$ from data will be presented.

\subsubsection{Local linear mapping}
\label{subsubsec: local linear mapping}

\begin{figure}
\centering
\includegraphics[scale=1]{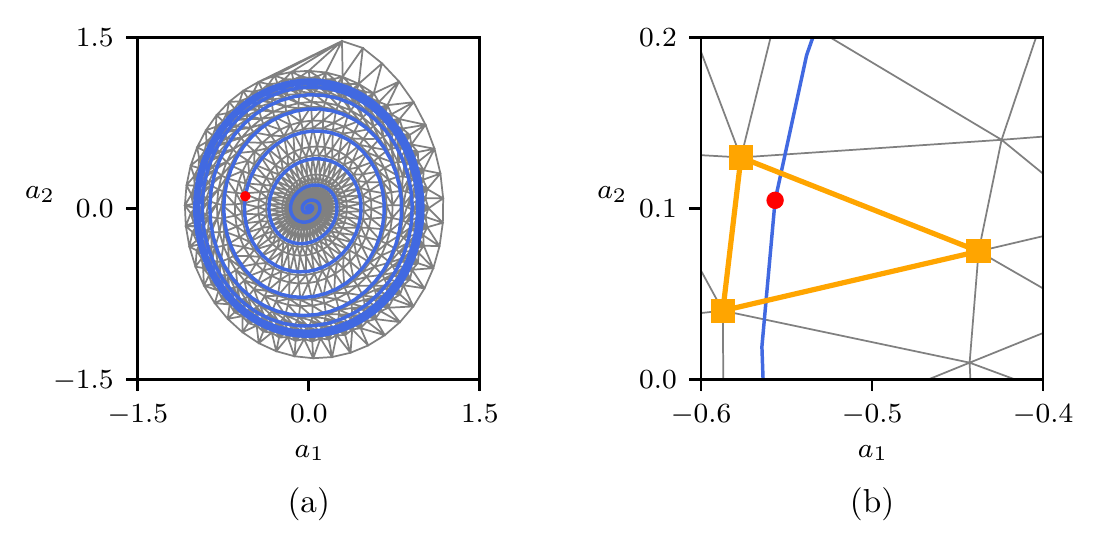}
\caption{In both figures, the blue line depicts the trajectory of the testing dataset for which we reconstruct the flow field. (a) Delaunay triangulation of the state plane. In addition to the two transient trajectories started from the fixed point, a third trajectory with an initial condition above the limit cycle has been used to obtain this triangulation. (b) Close-up view in the vicinity of the query point ({\color{red} $\bullet$}). The corresponding flow field can then be estimated as a weighted average of the flow fields associated to each vertex of the triangle highlighted in green.}
\label{fig: delaunay triangulation}
\end{figure}

Let us consider, for the sake of simplicity and without loss of generality, the feature vector $\bm{a} = \begin{bmatrix} a_1 & a_2 \end{bmatrix}^T$.
Given different transient evolutions of $\bm{a}(t)$, and having stored the associated velocity field snapshots, the nonlinear mapping $\bm{h}(\bm{x},\bm{a}) \approx \bm{u}(\bm{x}, t)$ can be approximated by a local linear mapping.
In the rest of this work, $\bm{a}^{\bullet}(t)$ will denote time-evolutions of the feature vector obtained from a direct numerical simulation, while $\bm{a}^{\circ}(t)$ will denote the evolution predicted by the low-dimensional models identified using the SINDy architecture outlined in \textsection \ref{subsec: SINDy}.
A Delaunay triangulation of the phase plane of the low-dimensional system can then be obtained from the transient evolutions of $a_1^{\bullet}(t)$ and $a_2^{\bullet}(t)$ in the training dataset.
An example triangulation is illustrated in figure \ref{fig: delaunay triangulation}(a).
Estimating the flow field associated with a point $\begin{bmatrix} a_1^{\circ} & a_2^{\circ} \end{bmatrix}$ then amounts to a two-step procedure:

\begin{enumerate}
\item Given the Delaunay triangulation of the phase plane, identify in which triangle the point $\begin{bmatrix} a_1^{\circ} & a_2^{\circ} \end{bmatrix}$ is contained.
See figure \ref{fig: delaunay triangulation} for an illustration.
\item Based on the vertices of this triangle, the flow field associated to
$\begin{bmatrix} a_1^{\circ} & a_2^{\circ} \end{bmatrix}$
can then be estimated as a weighted average of the flow fields associated to each vertex.
In the present work, these weights are chosen such that the query point
$\begin{bmatrix}a_1^{\circ} & a_2^{\circ}\end{bmatrix}$
is the barycentre of the corresponding triangle.
\end{enumerate}
Although it may be memory--intensive, since numerous snapshots need to be stored, it will be shown in \textsection \ref{subsec: flow field estimation} that this local linear mapping procedure allows for an unprecedented accuracy when reconstructing the flow field given only sensor measurements.
Further, it is possible to reduce the memory by compressing the snapshots.

\subsubsection{Feature-based modal expansion}\label{Sec:FeatureExpansion}
If full-state snapshots of the flow field are available, then it is possible to construct a set of generalized feature-based modes that are most correlated with either the feature vector ${\bm a}$ or with specific nonlinear terms in the sparse model \eqref{Eq:Dynamics}.
These feature-based modes make the sparse models physically interpretable, providing spatial structures associated with feature variables and specific interaction terms in the dynamics.

\setcounter{footnote}{1}
Consider a sequence of full-state snapshots\footnote{Note that the columns of the matrix ${\bm Q}$ correspond to a time-sequence of full-state snapshots, which is the mathematical convention in snapshot proper orthogonal decomposition and dynamic mode decomposition~\citep{Kutz2016book}.  However, the matrix ${\bm A}$ uses transposed notation to be consistent with the original SINDy paper~\citep{pnas:brunton:2016}, with rows corresponding to a time-sequence of transposed feature vectors ${\bm a}^T$. Thus, the columns of ${\bm A}^T$ are a time-sequence of the features ${\bm a}$, similar to ${\bm Q}$.}
\begin{align}
{\bm Q}({\bm x}) =
\begin{bmatrix}
{\bm u}^{m_1}({\bm x}) & {\bm u}^{m_2}({\bm x}) & {\bm u}^{m_p}({\bm x})
\end{bmatrix}.
\end{align}
Note that these snapshots do not need to be time-resolved, but they are collected at times $t^{m_1},\cdots,t^{m_p}$ that correspond to a subset of the resolved measurement times with which ${\bm s}$ and ${\bm a}$ are collected.
Recall that the columns of ${\bm A}^T$ are a time-resolved sequence of feature vectors ${\bm a}(t^m)$, with $m=1,\cdots,M$.
The columns of ${\bm A}^T$ associated with the non-time-resolved snapshots in ${\bm Q}$ are given by
\begin{align}
{\bm A}_{\bm Q}^T = \begin{bmatrix}{\bm a}(t^{m_1}) & {\bm a}(t^{m_2}) & \cdots & {\bm a}(t^{m_p}) \end{bmatrix}.
\end{align}
Thus, the snapshots sequence may be approximated by
\begin{align}
{\bm Q} = {\bm U}({\bm x}){\bm A}_{\bm Q}^T(t) + {\bm R}({\bm x}, t),
\end{align}
where ${\bm R}$ is the truncation residual. The columns of ${\bm U}({\bm x})$ are feature-based modes that are most correlated with the terms in the feature vector ${\bm a}(t)$. These modes are found via least-squares regression:
\begin{equation}
{\bm U} = {\bm Q} \left({\bm A}_{\bm Q}^T\right)^{\dagger},
\end{equation}
where $\left({\bm A}_{\bm Q}^T\right)^{\dagger}$ is the Moore-Penrose pseudo-inverse of ${\bm A}_{\bm Q}^T$.
In practice, this least-squares regression may be solved efficiently using the singular value decomposition of ${\bm A}_{\bm Q}^T$.

More generally, it is possible to compute modes that are most correlated with the dynamic interaction terms in the sparse model \eqref{Eq:SINDy1}.
Let $\gamma_1,\cdots,\gamma_q$ denote the indices of the rows in $\boldsymbol{\Xi}$ with non-zero entries, i.e. corresponding to active terms in the sparse dynamics.
The corresponding terms in the dynamics may be extracted via:
\begin{align}
\boldsymbol{\alpha} = \left(\boldsymbol{\Uptheta}({\bm a}) \begin{bmatrix}{\bm e}_{\gamma_1} & {\bm e}_{\gamma_2} & \cdots & {\bm e}_{\gamma_q} \end{bmatrix}\right)^T,
\end{align}
where ${\bm e}_{\gamma_j}$ is a column vector consisting entirely of zeros, except for a one in the $\gamma_j$-th row; i.e., ${\bm e}_{\gamma_j}$ is the $\gamma_j$-th column of the identity matrix.
For example, in the results, we will consider a vector of nonlinear terms $\boldsymbol{\alpha}$ given by $\boldsymbol{\alpha} = \begin{bmatrix} a_1 & a_2 & a_1^2 + a_2^2 & 2a_1a_2 & a_1^2 - a_2^2\end{bmatrix}^T$.
It is now possible to obtain generalized modes:
\begin{equation}
{\bm U} = {\bm Q} \left(\begin{bmatrix}\boldsymbol{\alpha}(t^{m_1}) & \boldsymbol{\alpha}(t^{m_2}) & \cdots & \boldsymbol{\alpha}(t^{m_p}) \end{bmatrix}\right)^{\dagger}.
\end{equation}
Thus, each mode ${\bm u}_i({\bm x})$ is a spatial field corresponding to a specific interaction term in the dynamical system, given by a component of $\boldsymbol{\alpha}$.

Compared to the local linear mapping presented in \textsection \ref{subsubsec: local linear mapping}, such a feature-based expansion has a low memory footprint, although it will typically be less accurate.
However, even if the feature-based modes are not used for full-state reconstruction, they imbue the sparse model with physical interpretability.
The modal representation above may be thought of as closely related to the proper orthogonal decomposition or dynamic mode decomposition, except generalized to identify modes that are most correlated with the features in ${\bm a}$ or the dynamic interaction terms in $\boldsymbol{\alpha}$.
à

%% file: S4_Results.tex
\section{Results}\label{section:results}

\subsection{Sensor-based dynamics}
\label{subsec: dynamical system}

In this section, a low-dimensional model of the transient and post-transient laminar cylinder wake is presented. First, a dynamical model capturing the dynamics of the lift coefficient is identified in \textsection \ref{subsubsec: lift-based gbm -- low-order model}. Then, in \textsection \ref{subsubsec: lift-and-drag based gbm -- descriptor system}, the low-order model aforementioned is supplemented with a nonlinear algebraic measurement equation in order to infer the evolution of the drag coefficient.

\subsubsection{Dynamical system}\label{subsubsec: lift-based gbm -- low-order model}

It is well known that the two-dimensional cylinder flow behaves as a self-excited, self-limiting and nearly harmonic nonlinear oscillator.
This behavior is clearly visible in the time-evolution of the instantaneous lift coefficient depicted in figure \ref{fig: lift signal}.
As such, the dynamics can be described by a nonlinear second-order ordinary differential equation, or by a set of two coupled first-order ODEs.
Given our feature vector ${\bm a}$ \eqref{eq: sensor measurement}, mapping the state of the system from the sensor-space to the feature-space, is defined as
\begin{equation}
{\bm a} = \begin{bmatrix}
				\hat{a}_1 & \hat{a}_2
			\end{bmatrix}^T,
\end{equation}
with $\hat{a} = \displaystyle \frac{a}{a_{\max}}$, $a_{\max}$ being the maximum value of $a$ once the system evolves onto the periodic limit cycle.
Such normalization ensures that
$$-1 \le a_{1, 2} \le 1,$$
a condition which greatly simplifies the sparse optimization problem involved in the identification procedure.
Although this mapping function has been defined analytically in the present work, similar features could be identified using delay coordinates, as in the singular spectrum analysis (SSA) in meteorology and ecology~\citep{ocean:colebrook:1978, geophys:barnett:1979, weathervec:weare:1982, geophys:ghil:2002}, NARMAX~\citep{Billings2013book}, the eigensystem realization algorithm (ERA) from system identification and control theory~\citep{jgcd:juang:1985}; these delay coordinates have recently been connected to the linear embedding of nonlinear dynamics via Koopman operator theory~\citep{Brunton2016havok,Arbabi2016arxiv}.
Alternatively, one could have also used the Hilbert transform for that purpose.

\begin{figure}
\centering
\includegraphics[scale=1]{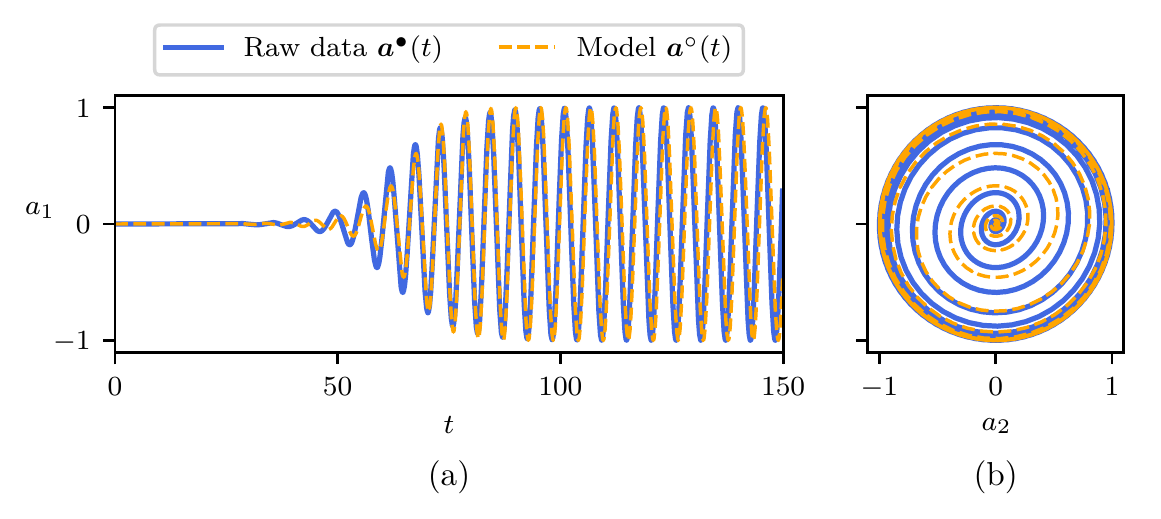}
\caption{(a) Comparison of the evolution as a function of time of the sensor measurement $a_1$ obtained from direct numerical simulation ({\color{blue} \bf -----} $a_1^{\bullet}$) and predicted by the identified low-dimensional model \eqref{eq: lift-based model} ({\color{orange} \bf -- --} $a_1^{\circ}$).
(b) Trajectory of the true and identified systems in the phase plane ($a_1$, $a_2$).
In both cases, the initial condition is close to the linearly unstable fixed point $\mathbf{u}_b(\mathbf{x})$, given by $\mathbf{a}=\mathbf{0}$.}
\label{fig: lift-based gbm - raw data vs model}
\end{figure}

Based on the different transient time evolutions of ${\bm s}^{\bullet}(t)$ recorded from direct numerical simulations, the corresponding feature vectors ${\bm a}^{\bullet}(t)$ have been computed and grouped into our training dataset.
The \emph{Sparse Identification of Nonlinear Dynamics} \citep{pnas:brunton:2016}, briefly outlined in \textsection \ref{subsec: SINDy}, is used to identify the equations governing the dynamics of ${\bm a}$.
The pool of candidate functions required for the identification process is chosen as
\begin{equation}
\Uptheta(a_1, a_2) = \begin{bmatrix} 1 & a_1 & a_2 & a_1^2 & a_1a_2 & a_2^2 & a_1^3 & a_1^2 a_2 & a_1 a_2^2 & a_2^3 \end{bmatrix}.
\label{eq: lift-based pool}
\end{equation}
Such a pool of polynomial functions, which can easily be enriched if needed, is a natural choice for the identification of a nonlinear oscillator~\citep{guckenheimer_holmes}.
In the present case, it leads to the identification of the following dynamical system
\begin{equation}
\displaystyle \frac{\mathrm{d}}{\mathrm{d}t}\begin{bmatrix}a_1 \\ a_2\end{bmatrix} = \begin{bmatrix} 0 & 1.12 \\ -1.116 & 0.28(1 - a_1^2 -a_2^2)\end{bmatrix}\begin{bmatrix}a_1 \\ a_2\end{bmatrix}.
\label{eq: lift-based model}
\end{equation}
More details about the model selection procedure can be found in appendix \ref{appendix: model selection}.
As shown in figure \ref{fig: lift-based gbm - raw data vs model}, the time evolution of $a_1^{\circ}(t)$ predicted by this low-dimensional system is in very good agreement with the one obtained for $a_1^{\bullet}(t)$ based on a direct numerical simulation whose initial condition has been chosen in the vicinity of the linearly unstable baseflow.
It is also remarkable that, despite its apparent simplicity, this two-degrees-of-freedom system captures all of the key physics of the cylinder flow, namely:
\begin{itemize}
\item It admits only one linearly unstable fixed point given by ${\bm a} = {\bm 0}$ and one attracting limit cycle characterized by $\Vert {\bm a} \Vert = 1$.
The corresponding circular frequency $\omega^{\circ} = 1.119$ in the nonlinearly saturated state is moreover less than 1.5\% smaller than the one observed in DNS ($\omega^{\bullet} = 1.132$).

\item It explicitly highlights the quadratic dependency of the instantaneous growth rate $2\sigma({\bm a}) =0.28(1 - a_1^2 - a_2^2)$.
Such quadratic dependencies are in line with our current understanding of the nonlinear saturation process of globally unstable flows; see \cite{prl:mantivc:2014} for more details.

\item Once the amplitude of the oscillation has saturated to $\Vert {\bm a} \Vert = 1$, the system reduces to a simple harmonic oscillator.
A similar structure could be derived based on a Galerkin projection of the Navier-Stokes equations onto the span of the first two POD modes and using the marginally stable mean flow as the reference state.
\end{itemize}
From a physical point of view, this low-order system describes the dynamics of the original high-dimensional system when constrained to the low-dimensional manifold structuring its phase phase \citep{jfm:noack:2003}.

\begin{figure}
\centering
\includegraphics[scale=1]{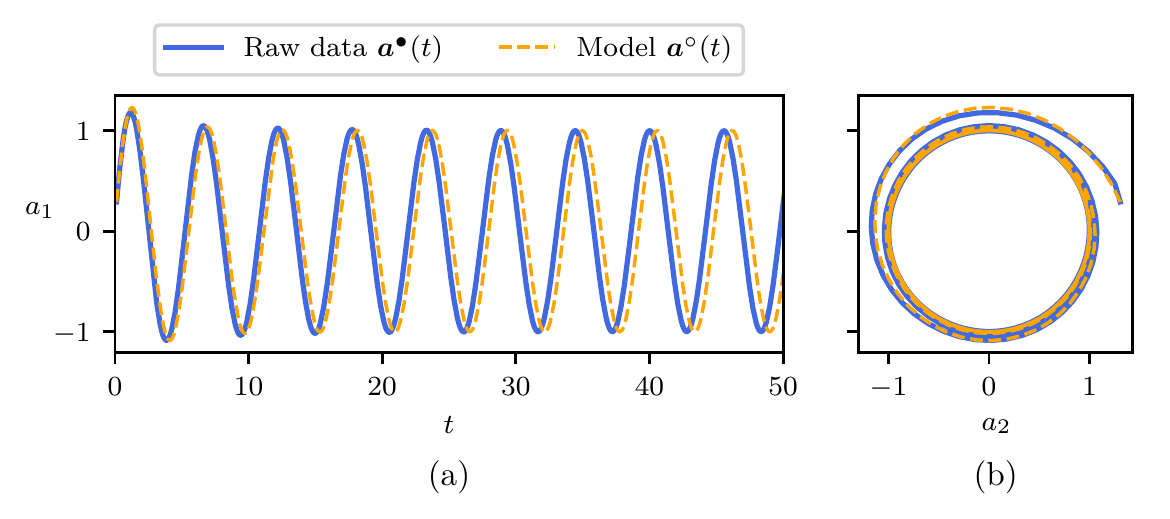}
\caption{(a) Comparison of the evolution as a function of time of the sensor measurement $a_1$ obtained from direct numerical simulation ({\color{blue} \bf -----} $a_1^{\bullet}$) and predicted by the identified low-dimensional model \eqref{eq: lift-based model} ({\color{orange} \bf -- --} $a_1^{\circ}$).
(b) Trajectory of the true and identified systems in the phase plane ($a_1$, $a_2$).
The initial condition used in the direct numerical simulation has been chosen to lie outside of the limit cycle, and for physical reasons it is also constrained to start close to the paraboloid manifold structuring the phase space of the system.}
\label{fig: lift-based gbm - raw data vs model - cross validation}
\end{figure}

Although the particular transient evolution of $a_1^{\bullet}(t)$ depicted in figure~\ref{fig: lift-based gbm - raw data vs model} has not been included in the training dataset, it is nonetheless characteristic of the type of evolution that our model has been specifically trained on.
One might then wonder how the model generalizes, for instance, given an initial condition that lies outside of the limit cycle.
Such an evolution is shown in figure \ref{fig: lift-based gbm - raw data vs model - cross validation} for an initial condition that has a radius from the fixed point that is 33\% larger than the radius of the limit cycle.
Note that, for physical reasons, the initial condition used in the direct numerical simulation has been chosen such that it is close to the low-dimensional inertial manifold.
Once again, the dynamics of $a_1^{\circ}(t)$ predicted by \eqref{eq: lift-based model} are in qualitative and quantitative agreements with the actual dynamics of $a_1^{\bullet}(t)$ obtained from the corresponding direct numerical simulation.
Finally, it is worth noting that, although we have aimed for a continuous-time representation of the dynamics in this work, the present methodology can easily be adapted to obtain a discrete-time version of the same system, as described in appendix~\ref{appendix: discrete-time}.

\subsubsection{Descriptor system}\label{subsubsec: lift-and-drag based gbm -- descriptor system}

The dynamical system identified in \textsection \ref{subsec: dynamical system} provides valuable information and accurate predictions of the dynamics of the system.
However, being based solely on quantities derived from the instantaneous lift coefficient, the model does not enable direct estimation of the corresponding drag force.
Having a model that can estimate the instantaneous drag force might, however, be of critical importance in aerodynamic applications.
Defining the drag coefficient as a degree of freedom of our system, two different possibilities are then available: (i) extend the dynamical system \eqref{eq: lift-based model} with a (nonlinear) algebraic measurement equation, or (ii) identify a new dynamical system made of three coupled ODEs.
We have investigated both of these options, and found that models may be successfully identified.
Given the simplicity and accuracy of the dynamical system \eqref{eq: lift-based model} identified in \textsection \ref{subsec: dynamical system}.1, the former approach is preferred over the latter.
For that purpose, an additional feature is added to ${\bm a}$, given by
\begin{equation}
a_3(t) = \frac{C_D(t) - C_{D_0}}{\max(C_D(t) - C_{D_0})},
\notag
\end{equation}
where $C_{D_0}$ denotes the instantaneous drag coefficient of the base flow ${\bm u}_s({\bm x})$, and $\max(C_D(t) - C_{D_0})$ is the maximum value once the flow has reached its statistically stationary state.
We thus seek a nonlinear algebraic measurement equation of the form
\begin{equation}
a_3 = f(a_1, a_2).
\notag
\end{equation}
This algebraic measurement equation can also be identified using sparse regression with the same pool $\Uptheta(a_1, a_2)$ of candidate functions \eqref{eq: lift-based pool} as before.
Given our training dataset, the parsimonious measurement equation identified reads
\begin{equation}
a_3^2 = 0.97 a_1^2 - 0.16 a_1 a_2 + 0.84 a_2^2.
\label{eq: measurement equation}
\end{equation}
Combining this measurement equation with the dynamical system \eqref{eq: lift-based model} identified in \textsection \ref{subsubsec: lift-based gbm -- low-order model} results in a low-dimensional descriptor system governing the evolution of both the instantaneous lift and drag coefficients.

Equation \eqref{eq: measurement equation} describes a distorted cone and is notably different from the mean-field paraboloid of the 3-dimensional POD-Galerkin model by \cite{jfm:noack:2003}. This difference can be explained by the choice of the state space $\bm{a}$. The oscillation of von K\'arm\'an vortex shedding is characterized by $a_1$ and $a_2$---albeit from different physical mechanisms. During the transient, the vortex shedding moves upstream from the stagnation point of the unstable steady solution
to the immediate vicinity of the cylinder. The global POD mode amplitudes $a_1$, $a_2$ resolve the vortex shedding already far downstream, while the lift-based feature coordinates 'feel' the vortex shedding only in the final stage when the vortices 'rub' on the cylinder. In contrast, the drag variation and the shift-mode amplitude of the 3-dimensional POD-Galerkin model are linearly related.
Thus, $a_3$ of the Galerkin expansion and of the force-related ROM resolve the same physics.
Hence, the fluctuation amplitude $\sqrt{a_1^2+a_2^2}$ increases much faster with $a_3$  for the POD-based model than for the force-based system.

\begin{figure}
\centering
\includegraphics[scale=1]{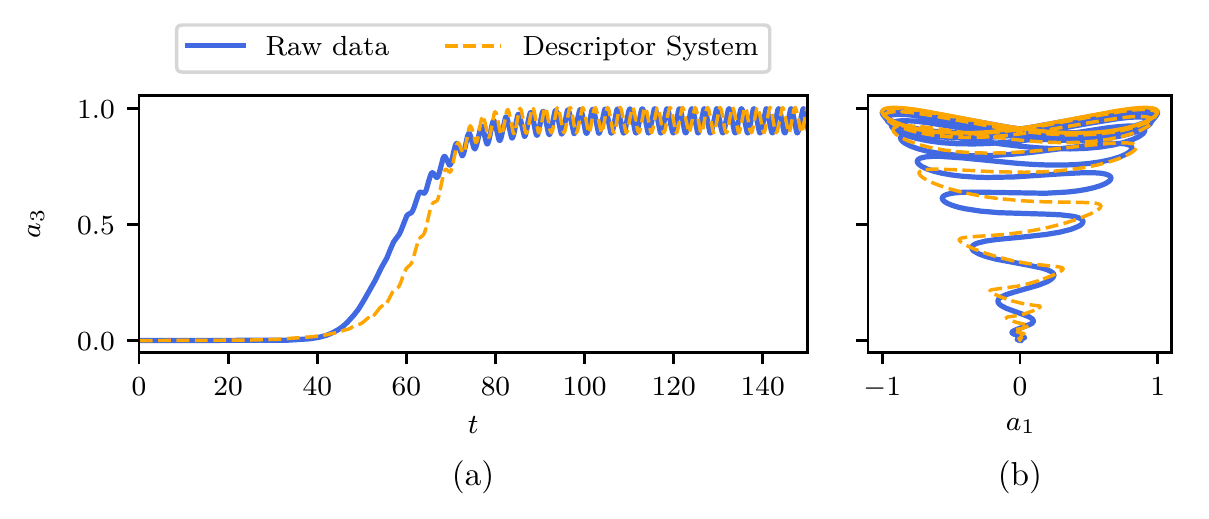}
\caption{(a) Comparison of the evolution as a function of time of the sensor measurement $a_3$ obtained from direct numerical simulation ({\color{blue} \bf -----} $a_3^{\bullet}$) and predicted by the identified low-dimensional descriptor system made of \eqref{eq: lift-based model} and \eqref{eq: measurement equation} ({\color{orange} \bf -- --} $a_3^{\circ}$).
(b) Trajectory of the true system and the identified one in the phase plane ($a_1$, $a_3$).}
\label{fig: descriptor system - raw data vs model}
\end{figure}

\begin{figure}
\centering
\includegraphics[scale=1]{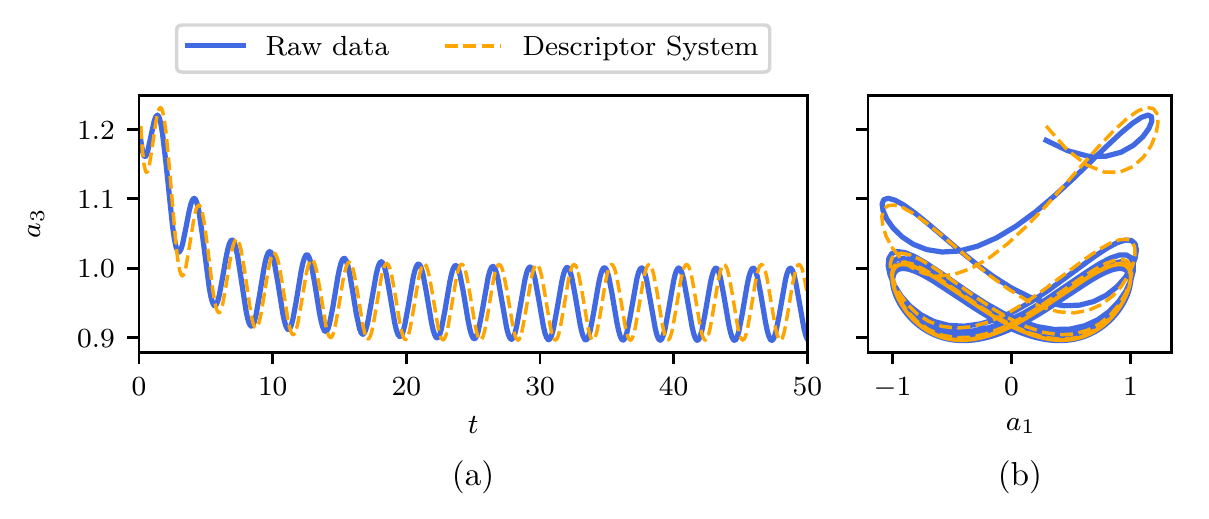}
\caption{(a) Comparison of the evolution as a function of time of the sensor measurement $a_3$ obtained from direct numerical simulation ({\color{blue} \bf -----} $a_3^{\bullet}$) and predicted by the identified low-dimensional descriptor system made of \eqref{eq: lift-based model} and \eqref{eq: measurement equation} ({\color{orange} \bf -- --} $a_3^{\circ}$).
(b) Trajectory of the true system and the identified one in the phase plane ($a_1$, $a_3$).}
\label{fig: descriptor system - raw data vs model cross-validation}
\end{figure}

Figure \ref{fig: descriptor system - raw data vs model}(a)  and figure \ref{fig: descriptor system - raw data vs model cross-validation}(a) compare the predicted time-evolution of $a_3^{\circ}(t)$ against the evolution of $a_3^{\bullet}(t)$ obtained from direct numerical simulation, whereas figure \ref{fig: descriptor system - raw data vs model}(b) and figure \ref{fig: descriptor system - raw data vs model cross-validation}(b) depict the associated trajectories projected onto the ($a_1$, $a_3$) phase plane.
Along with the dynamics of $a_1$ and $a_2$ being correctly captured by the dynamical system \eqref{eq: lift-based model}, it is clear that the nonlinear algebraic measurement equation \eqref{eq: measurement equation} correctly infers the evolution of $a_3$.
Although a small misprediction of the amplitude of $a_3$ can be seen for $45 \le t \le 65$, the low-dimensional descriptor system identified in the present work is nonetheless one of the simplest and yet most accurate and physically interpretable low-order models available in the literature to reproduce the dynamics of the cylinder flow at $\Rey=100$.

\subsection{Flow field estimation}
\label{subsec: flow field estimation}

The descriptor system identified in the previous section provides valuable information and accurate predictions of the dynamics of the system.
However, it does not allow us to directly infer what the corresponding flow field is.
In order to estimate the flow, we thus need to supplement our dynamical system with a full-state estimator given by
\begin{equation}
{\bm u}({\bm x}, t) \approx {\bm h}({\bm x}, {\bm a}(t)).
\notag
\end{equation}
Formally, this full-state estimator ${\bm h}$ is a nonlinear mapping from the low-dimensional feature-space to the high-dimensional physical space.
In the rest of this section, two different strategies will be employed in order to build this nonlinear mapping: the local linear mapping procedure described in \textsection \ref{subsec: full-state estimation}, or a feature-based modal expansion of the velocity field in terms of the feature modes introduced in \textsection \ref{Sec:FeatureExpansion}.

Given a sparse nonlinear model, it is also possible to construct a generalized feature-based modal decomposition that identifies modal structures that are most correlated with specific terms in the dynamics.
We define a new feature vector $\boldsymbol{\alpha}$ containing all of the nonzero terms identified in the right hand side of the sparse model:
\begin{equation}
\boldsymbol{\alpha} \triangleq \begin{bmatrix}
				a_1 \\
				a_2 \\
				a_1^2 + a_2^2 \\
				2a_1 a_2 \\
				a_1^2 - a_2^2
			\end{bmatrix} = \begin{bmatrix}
				\alpha_1 \\
				\alpha_2 \\
				\alpha_3 \\
				\alpha_4 \\
				\alpha_5
			\end{bmatrix}.
\end{equation}
The feature-based modal expansion considered hereafter then reads
\begin{equation}
{\bm u}({\bm x}, t) = {\bm u}_0({\bm x}) + \sum_{i=1}^5 {\bm u}_i({\bm x}) \alpha_i(t) + {\bm r}({\bm x}, t)
\label{eq: flow estimation -- galerkin expansion}
\end{equation}
where ${\bm u}_0({\bm x})$ is the linearly unstable steady solution to the Navier-Stokes equations and ${\bm r}({\bm x}, t)$ the residual.
The different feature modes ${\bm u}_i({\bm x})$ have been computed following the procedure described in \textsection \ref{subsec: full-state estimation}.
Mathematically, the generalized mode decomposition is achieved with a simple least-squares regression:
\begin{align}\label{Eq:GMD}
{\bm Q} &\approx  \underbrace{\begin{bmatrix} \vline & \vline & \vline & \vline & \vline \\ {\bm u}_1({\bm x}) &  {\bm u}_2({\bm x}) &  {\bm u}_3({\bm x}) &  {\bm u}_4({\bm x}) &  {\bm u}_5({\bm x}) \\ \vline & \vline & \vline & \vline & \vline\end{bmatrix}}_{\textbf{Modes: } {\bm U} }
\underbrace{\begin{bmatrix} \rule[.5ex]{2.4em}{0.4pt}  ~\alpha_1(t)~  \rule[.5ex]{2.4em}{0.4pt}\\
 \rule[.5ex]{2.4em}{0.4pt}  ~\alpha_2(t)~  \rule[.5ex]{2.4em}{0.4pt}\\
  \rule[.5ex]{2.4em}{0.4pt}  ~\alpha_3(t)~  \rule[.5ex]{2.4em}{0.4pt}\\
   \rule[.5ex]{2.4em}{0.4pt}  ~\alpha_4(t)~  \rule[.5ex]{2.4em}{0.4pt}\\
    \rule[.5ex]{2.4em}{0.4pt}  ~\alpha_5(t)~  \rule[.5ex]{2.4em}{0.4pt} \end{bmatrix}}_{\textbf{Time dynamics:  }\boldsymbol{\alpha}(t)}
\quad\Longrightarrow\quad {\bm U} = \bm{Q}\left( \boldsymbol{\alpha}(\mathbf{t})\right)^\dagger
\end{align}
where $\left(\boldsymbol{\alpha}({\bm t})\right)^\dagger$ denotes the pseudo-inverse of the matrix $\boldsymbol{\alpha}({\bm t})$, and $\bm{Q}$ is the sequence of baseflow-substracted snapshots.

\begin{figure}
\centering
\includegraphics[scale=1]{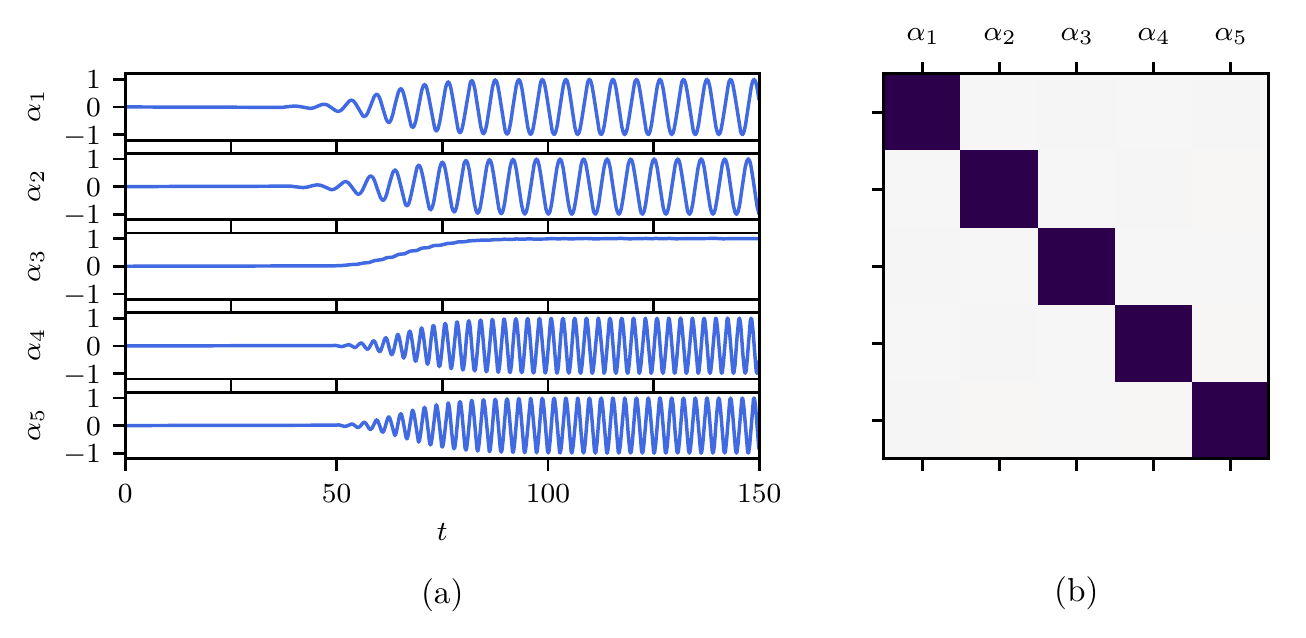}
\caption{(a) Time evolution of the different basis coefficients used in the feature-based modal expansion~\eqref{eq: flow estimation -- galerkin expansion}.
(b) Corresponding cross-correlation matrix. Dark squares indicate that $\alpha_i$ and $\alpha_j$ are strongly correlated, while white squares indicate they are uncorrelated.}
\label{fig: flow estimation-- cross correlation matrix}
\end{figure}

\begin{figure}
\centering
\includegraphics[scale=1]{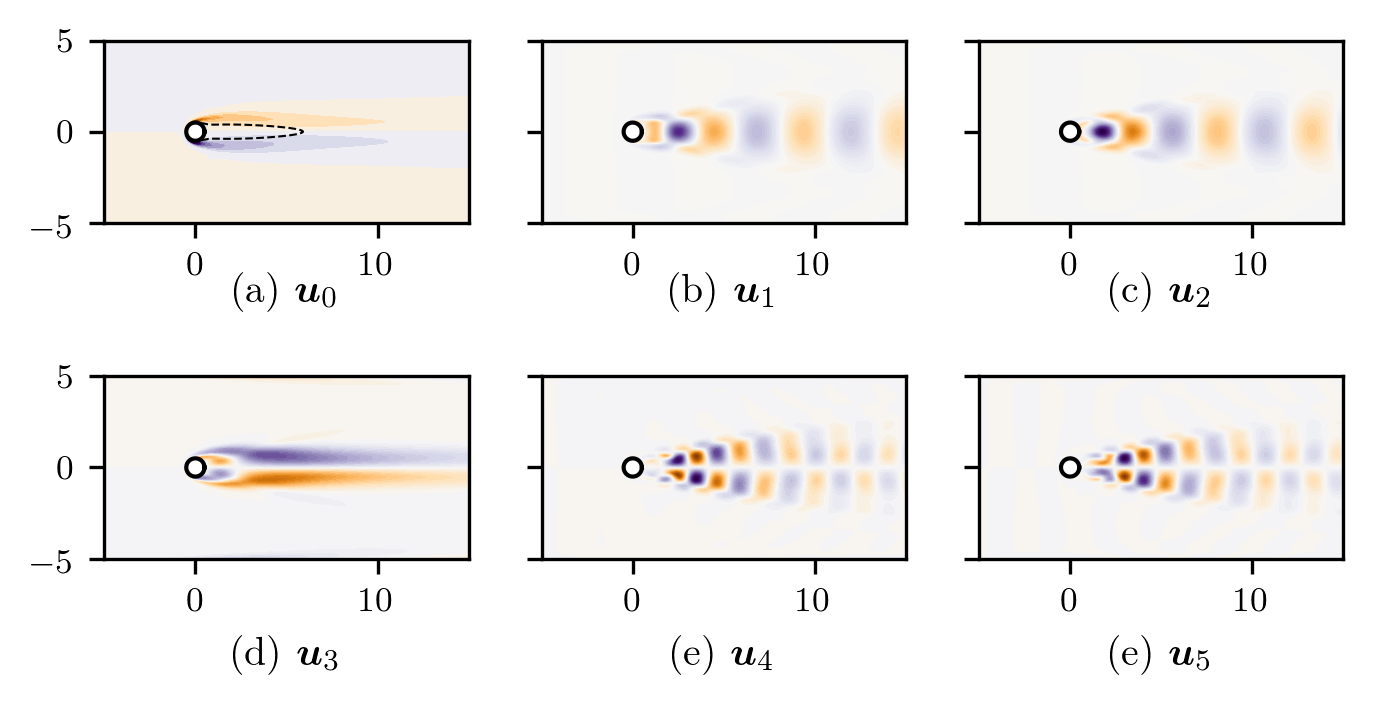}
\caption{Figure (a) depicts the vorticity field of the linearly unstable base flow, while the dashed line highlights the spatial extent of the reversed flow region. Figurs (b) to (e) show the vorticity field of the feature modes associated to the feature vector ${\bm b}$.}
\label{fig: flow estimation -- feature modes}
\end{figure}

The associated vorticity fields are shown in figure \ref{fig: flow estimation -- feature modes}, while the time-evolution of the different basis coefficients $\alpha_i(t)$ is depicted in figure \ref{fig: flow estimation-- cross correlation matrix}(a). Figure \ref{fig: flow estimation-- cross correlation matrix}(b) shows the cross correlation matrix of these signals. Given its diagonal structure, it is clear that the different basis coefficients $\alpha_i(t)$ are uncorrelated one to another.
For the cylinder flow, these feature modes are very similar to the classical POD modes.
A key advantage over POD modes is that the present modes are directly interpretable as being the coherent structures most correlated with our different measurements and the sparse nonlinear interaction terms in the model.
Moreover, the shift mode ${\bm u}_3$ naturally arises in this framework as the result of quadratic interactions between $a_1$ and $a_2$, consistent with our understanding of the nonlinear saturation of globally unstable flows \citep{prl:mantivc:2014}.
Defining the feature vector $\boldsymbol{\alpha}$ used in the feature-based modal expansion such that it includes quadratic terms was thus deemed necessary in order to ensure that the distortion between the linearly unstable base flow and the marginally stable mean flow is correctly captured by the flow estimator using the feature-based modal expansion~\eqref{eq: flow estimation -- galerkin expansion}.

\begin{figure}
\centering
\includegraphics[scale=1]{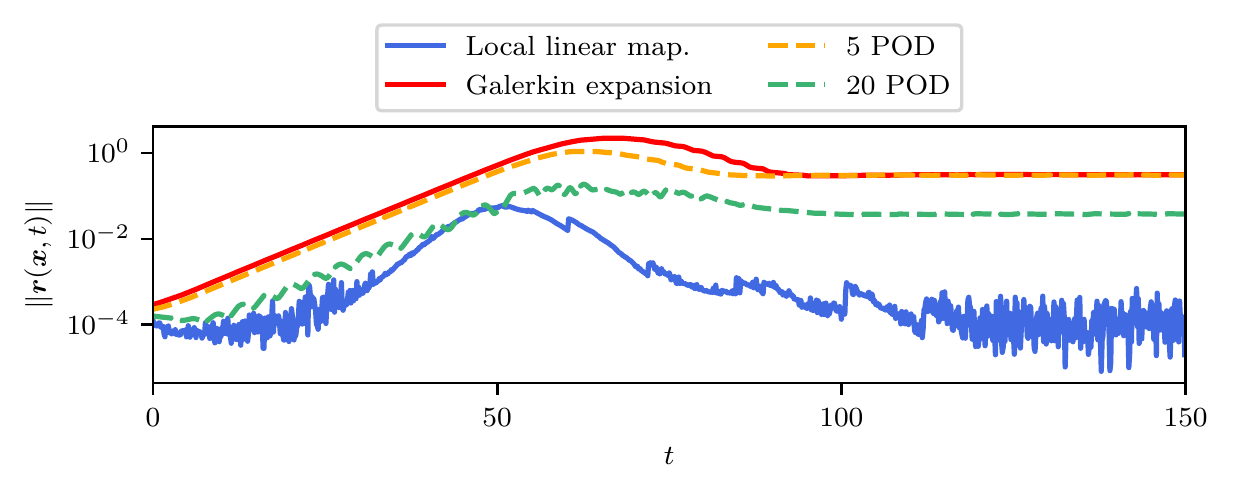}
\caption{Time evolution of the estimation error $\Vert {\bm r}(\bm{x}, t) \Vert$ for different flow estimators.}
\label{fig: flow estimation -- residual evolution}
\end{figure}

Figure \ref{fig: flow estimation -- residual evolution} compares the evolution of $\Vert {\bm r} \Vert$, i.e.\ the norm of the estimation error, for the local linear mapping and the feature-based modal expansion \eqref{eq: flow estimation -- galerkin expansion}.
The evolution of the truncation error for different POD bases is also reported for the sake of comparison.
The flow estimator based on the local linear mapping (LLM) largely outperforms the two estimators based on a 5-mode POD expansion and the generalized modal decomposition with 5 modes.
Its very good performances, on average two to three orders of magnitude more accurate than the other estimators considered, results from the fact that LLM leverages all of the information contained in the different snapshots matrices used whereas the different modal expansions only provide low-rank approximations of these same matrices.
However, this higher accuracy comes at the price of increased storage.
This requirement can nonetheless be mitigated by computing a low-rank approximation of the snapshot matrix used in the training dataset.
In the present case, considering a rank-50 approximation based on the singular value decomposition has almost no effect on the estimation error, while significantly reducing the memory requirements.
Instead of storing $M=1,200$ snapshots, each a 100,000-dimensional vector corresponding to a vertex of the Delaunay triangulation of phase space, only the 50 leading full-state singular vectors must be stored, along with the $50$-dimensional vector of coefficients for each vertex.
This reduces the storage requirements by a factor of nearly $24$.
In this case, the local linear mapping would then approximate the transfer function mapping the state of our system from the 2-dimensional phase space of the dynamical system \eqref{eq: lift-based model} to the 50-dimensional space of coefficients resulting from the truncated singular value decomposition of the snapshots matrix.

\begin{figure}
\centering
\includegraphics[scale=1]{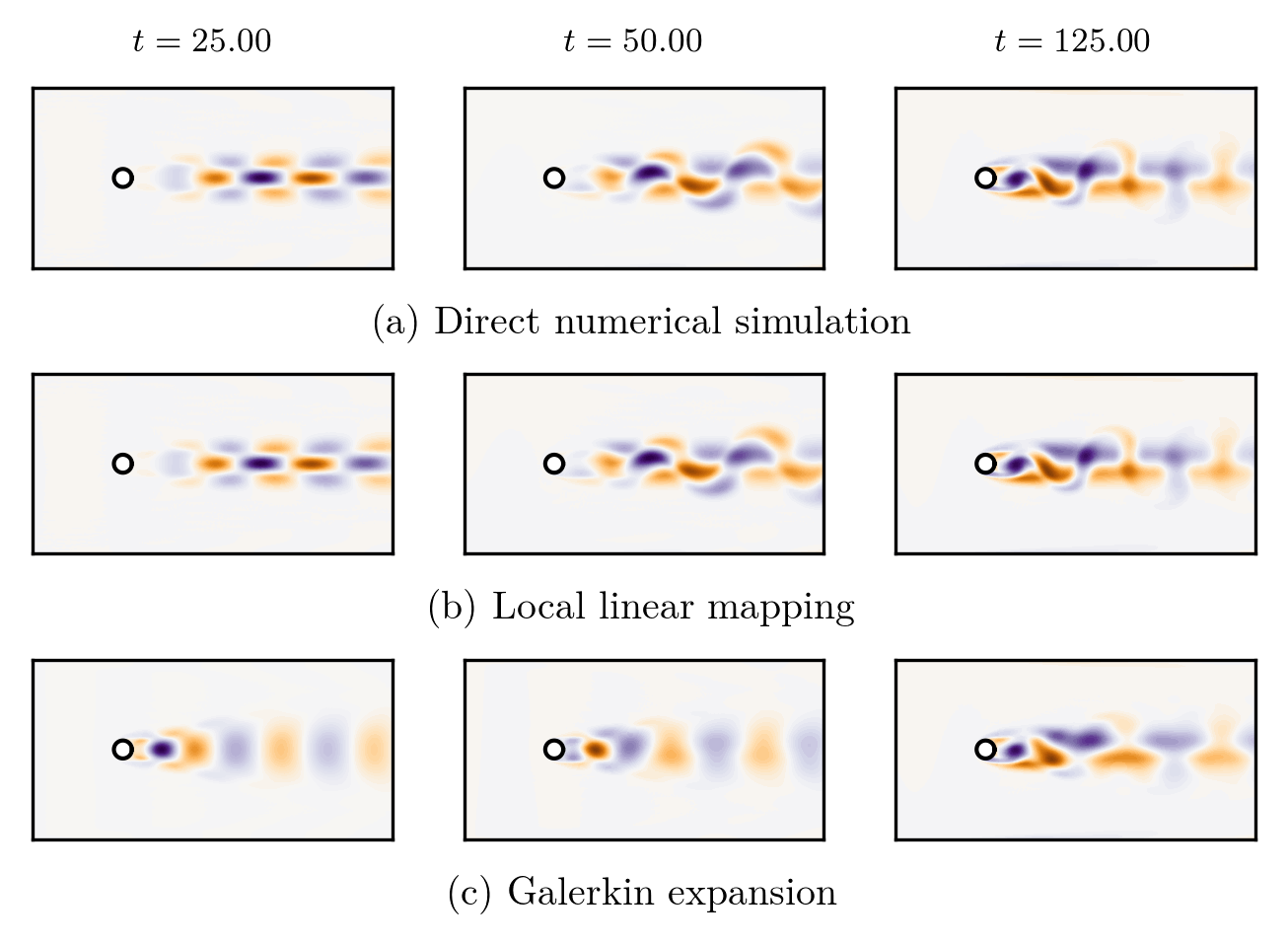}
\caption{Comparison of the vorticity fields at different instants of time obtained (a) from direct numerical simulation, (b) from the local linear mapping procedure and (c) from the feature-based modal expansion.
Note that the vorticity field of the linearly unstable baseflow has been subtracted in order to highlight the perturbative vorticity field.}
\label{fig: flow estimation -- estimated flow fields}
\end{figure}

Finally, figures \ref{fig: flow estimation -- estimated flow fields}(b) and (c) depict the estimated vorticity field, with the base flow subtracted, at different instants in time and compare them with the true vorticity field obtained from direct numerical simulation in figure \ref{fig: flow estimation -- estimated flow fields}(a).
The vorticity fields estimated using the local linear mapping are in much better agreement, from a physical and kinematic point of view, than the ones obtained by the 5-mode modal expansion.
This is especially pronounced during the period of exponential growth of the linear instability and at the onset of nonlinear saturation during which the local linear mapping correctly captures the deformation and distortion of the flow structure.
In contrast, low-rank modal expansions are notorious for their inability to capture such mode deformation and/or changes in operating conditions.
Given that the POD modes and generalized feature modes used in this work essentially approximate the flow structure once the system has reached the periodic limit cycle, it is thus expected that they provide only a very crude estimation of the flow structures when the system evolves in the vicinity of the linearly unstable base flow.
This inability of a modal expansion to easily address mode deformation is one of its key limitations and is the principal reason why the local linear mapping strategy should be preferred.
Combining the descriptor system identified in \textsection \ref{subsec: dynamical system} with the local linear flow estimator finally allows us to construct a two-degrees-of-freedom reduced-order model of the cylinder flow, having an unprecedented accuracy.

%% file: S5_Conclusions.tex
\section{Conclusions}\label{section:conclusions}

This work develops a new reduced-order modeling procedure for unsteady fluid flows that yields accurate nonlinear models and insight into relevant flow structures.
This procedure identifies sparse nonlinear models, not on the full fluid state, but from time-resolved sensor measurements that may be realistically obtained in experiments.
The sparsity of the model prevents overfitting and uncovers key nonlinear interaction terms.
If PIV snapshots are also available, not necessarily time-resolved, it is possible to estimate the full-state from the sparse model using local linear mapping: the full-state is interpolated between the most similar historical flow fields, based on the dynamics.
It is also possible to construct a generalized modal decomposition that identifies coherent structures most correlated with each interaction term in the sparse nonlinear model.

Our methodology, summarized in figure \ref{Fig:overview}, can be divided into four steps:
\begin{enumerate}
	\item Given a set of physically relevant sensor measurements ${\bm s}$, create a low-dimensional feature vector ${\bm a}$ that allows for a complete dynamics characterization of the state of the system. Although we have used an analytical mapping ${\bm a} = {\bm g}({\bm s})$ in the present work, this procedure could also be performed, for instance, by means of kernel PCA~\citep{Scholkopf1998nc}.
	Moreover, the sensors could come from force measurements, as in the present work, or from any other time resolved sensor measurements, such as pressure along the body.
	In the case of limited measurements, it may also be possible to augment the state using delay coordinates~\citep{takens:1981}, in the spirit of SSA~\citep{ocean:colebrook:1978, geophys:barnett:1979}, NARMAX~\citep{Billings2013book}, ERA~\citep{jgcd:juang:1985}, or HAVOK~\citep{Brunton2016havok} models; delay coordinates have since been connected to Koopman operator theory~\citep{Brunton2016havok,Arbabi2016arxiv}.

	\item The second step identifies the equations governing the dynamics of ${\bm a}$. In this study, the SINDy algorithm~\citep{pnas:brunton:2016} yields a sparse nonlinear model in continuous-time.
	Discrete-time models may also be obtained via SINDy, or using the classical NARMAX or Volterra series approaches \citep{Semeraro2016arxiv}.
	In the past, nonlinear reduced-order models have primarily been obtained with knowledge of the governing equations and with the use of high-fidelity numerical simulations, such as in the Galerkin projection procedure.
	The present approach is exclusively data-driven, and does not rely on prior knowledge of the governing equations, which is consistent with the modern machine learning approach.
	Thus, interpretable models may be obtained from experimental measurements alone.

	\item If full-state snapshots are available, it has been shown in \textsection \ref{subsubsec: local linear mapping} and \textsection \ref{subsec: flow field estimation} that the low-order model can be supplemented with a full-state estimator ${\bm u}({\bm x}, t) \approx {\bm h}({\bm x},{\bm a})$.
	The full-state snapshots do not need to be time-resolved, but they do need to be synchronized with the sensor measurements.
	The local linear mapping procedure provides a highly accurate approximation ${\bm h}({\bm x},{\bm a})$.
	Full-state estimation is particularly relevant in control  applications~\citep{fabbiane2014amr,Brunton2015amr,Sipp2016amr}, where feedback control is often designed on the full state, which is then estimated from sensor measurements~\citep{dp:book,sp:book}.
	If the sensor measurements are noisy, it may also be helpful to design a Kalman filter~\citep{Kalman1960jfe,Welch1995book} based on the sparse dynamics, so that the estimated state does not jump in response to noise jitter.

	\item Although the local linear mapping yields the most accurate full-state reconstruction, it may also be useful to identify a generalized modal expansion for interpretability of the low-order model.
	Through simple regression, it is possible to identify modal structures that are most correlated with each term in the SINDy model.
	This procedure is similar in concept to DMD, but generalized to identify structures associated with nonlinear terms in the model.
	These coherent structures may improve intuition and interpretability.
Some challenges and improvements of DMD for transient cylinder wakes have been described by \citet{Noack2016jfm}.
\end{enumerate}

This methodology is illustrated using the canonical two-dimensional cylinder flow at $\Rey = 100$.
Despite its simplicity, this flow configuration is a prototypical example capturing the key physics of bluff body flows.
Various models are identified based on sensor measurements of the lift and drag coefficients, $C_L$ and $C_D$, which are physically relevant and are readily accessible in experiments.
First, a dynamical system~\eqref{eq: lift-based model} is identified using a single sensor input $s = C_L$, which is augmented with its time derivative. As discussed in \textsection \ref{subsubsec: lift-based gbm -- low-order model}, this system models the cylinder flow dynamics, which evolve along a low-dimensional manifold.
Next, in \textsection \ref{subsubsec: lift-and-drag based gbm -- descriptor system}, the same dynamical system may be supplemented with a sparse nonlinear equation to build an algebraic representation of the drag coefficient $C_D$ in terms of the evolution of $C_L$.
It is also possible to build a reduced model with three degrees of freedom in terms of $C_L$, its time derivative, and $C_D$.
This model is the most general and has a broader range of validity, as it captures the low-dimensional slow-manifold structure underlying phase space.
Finally, the local linear mapping in \textsection \ref{subsec: flow field estimation} provides full-state estimation with unprecedented accuracy compared to the classical Galerkin expansion.

Even though this study uses data from direct numerical simulations, the overall strategy is generally applicable to a real flow experiment with minor modifications. Despite their simplicity, the identified models do not suffer the same drawbacks as reduced-order models obtained from a Galerkin projection procedure, namely over-estimation of the duration of transients and energy overshoots at the onset of nonlinear saturation. Instead, the identified sparse models provide simple explanations for the nonlinear saturation process of globally unstable flows, as in \eqref{eq: lift-based model}. Moreover, the models are based on sensor measurements, which may include lift, drag, or pressure measurements that are physically linked to the geometry. Working in these \emph{intrinsic} coordinates has the potential to overcome many of the limitations of classical modal-based projection methods, including mode deformation due to moving geometry and varying parameters.

The effectiveness of the reduced-order models identified and the modularity of the methodology proposed in the present work suggest a number of exciting future directions.
There is significant potential for these methods to be applied broadly to obtain interpretable reduced-order models for a range of flow configurations in simulations and experiments.
For example, these sensor-based models may be applied to develop nonlinear unsteady aerodynamic models, generalizing previous linear and linear parameter varying models~\citep{Brunton2013jfm,Brunton2014jfs,Hemati2016aiaa}.

A key perspective to be given to this work is its extension to flow control.
Given a feature vector ${\bm a}$ and actuators characterized by a control law ${\bm b}(t)$, one could use SINDy with control (SINDYc)~\citep{Brunton2016nolcos} in order to identify low-order models

\begin{equation}
	\frac{\mathrm{d} {\bm a}}{\mathrm{d}t} = {\bm f}({\bm a}, {\bm b})
	\notag
\end{equation}

\noindent that incorporate the influence of the actuation ${\bm b}$ on the state ${\bm a}$.
Combining such an approach with \emph{Machine Learning Control} \citep{duriez2016machine} may result in interpretable models of entirely new flow behaviors and previously unobserved flow physics that are discovered through in the controlled flow.
The identified models can then serve as a low-dimensional representation of the actual system in order to facilitate the computation of nonlinear optimal feedback control laws.
This is an area of active research by the authors.
In the near future, the authors aim to apply the methodology introduced in the present work to the optimal control of experimental flows.

Finally, there are a number of methodological extensions that may improve the performance of this sparse modeling framework.
First, it will be important to demonstrate that these methods scale favorably to systems with higher-dimensional attractors; however, because the algorithms are based on simple regression and sparse optimization, they should remain computationally tractable.
Next, it may be possible to reduce the memory requirements of the local linear mapping by building local modal libraries in different dynamic regimes (e.g., linear instability, saturated limit cycle, etc.).
The storage requirements may further be reduced using compression techniques and sparse sampling.
Finally, it may be possible to incorporate the accuracy of the generalized modal decomposition reconstruction into the cost function in the SINDy regression, so that nonlinear features are selected based on their dynamic relevance and their ability to correlate to full-state structures.

\section*{Acknowledgments}

SLB acknowledges generous funding support from the Defense Advanced Research Projects Agency (DARPA HR0011-16-C-0016) and from the Air Force Office of Scientific Research (AFOSR FA9550-13-1-0183).
SLB would like to thank Nathan Kutz, Josh Proctor, Niall Mangan, and Sam Rudy for discussions related to sparse model identification.
SLB would also like to thank Scott Dawson for valuable discussions related to nonlinear modeling in terms of aerodynamic force coefficients.

BRN acknowledges the funding and excellent working conditions
of the Collaborative Research Center (CRC880)
'Fundamentals of High Lift for Future Civil Aircraft'
funded by the German Science Foundation (DFG)
and hosted by the Technical University of Braunschweig, Germany.
This work is also supported by internal funds of LIMSI-CNRS
and a public grant overseen
by the French National Research Agency (ANR) as part of the ``Investissement d’Avenir'' program,
through the  ``iCODE Institute project'' funded by the IDEX Paris-Saclay, ANR-11-IDEX-0003-02.

%% file: SA_Appendices.tex
\section{Model Selection}
\label{appendix: model selection}

Model selection and cross-validation are crucial components of system identification, as regression models tend to overfit with increasing training data, unless care is taken. The goal is to identify, among all candidate models, the parsimonious model that optimally balances model accuracy and model complexity.
As the sparsifying parameter $\lambda$ is varied in the SINDy procedure, a Pareto front is swept out, reducing the combinatorially many candidate models down to a small handful of candidates models.
\cite{arxiv:mangan:2016} have recently demonstrated how SINDy can be combined with the well-known Akaike information criterion (AIC)~\citep{Akaike1974ieeetac} or the Bayes information criterion (BIC)~\citep{Schwarz1978BIC} in order to select the most parsimonious model from this Pareto front.
Given a candidate model, the associated AIC score is given by
\begin{equation}
AIC = 2k - 2 \ln(L({\bm a}, \mu)) + 2\frac{(k+1) (k+2))}{(m-k-2)},
\label{eq: AIC score}
\end{equation}
where $L({\bm a}, \mu)$ is the loss function of the observations ${\bm a}$ given the best-fit parameters values $\mu$ of the candidate model and $k$ the total number of free parameters. The last term in \eqref{eq: AIC score} is a finite sample size correction where $m$ is the total number of observations used to cross-validate the model. For two models of the same accuracy, the AIC score will penalize the one having the larger number of free parameters. In this work, the loss function has been chosen as
\begin{equation}
L({\bm a}, \mu) = \frac{1}{m}\sum \frac{\int_0^T \Vert {\bm a}^{\circ}(t) - {\bm a}^{\bullet}(t) \Vert^2 \mathrm{d}t}{\int_0^T  \Vert {\bm a}^{\bullet}(t) \Vert^2 \mathrm{d}t}
\end{equation}
where ${\bm a}^{\bullet}(t)$ is the time-evolution obtained from direct numerical simulation of the Navier-Stokes equations and ${\bm a}^{\circ}(t)$ is the evolution predicted by the low-dimensional model considered. The summation is over the $m$ different training and testing datasets used for cross-validation. The AIC scores for each candidate model can have a wide range of values, hence requiring a rescaling by the minimum AIC value. The relative AIC score is thus given by
\begin{equation}
\Delta = AIC -AIC_{\min}.
\end{equation}
The different candidate models can then be ranked based on this relative AIC score. Following \cite{arxiv:mangan:2016}, models with $\Delta \le 2$ have so-called \emph{strong support}, models with $4 \le \Delta \le 7$ have \emph{weak support}, and models with $\Delta \ge 10$ have \emph{no support}. It should be emphasized that the model characterized by $\Delta = 0$ is not necessarily the best model possible, but only the best one among the different models tested.

\begin{figure}
\centering
\includegraphics[scale=1]{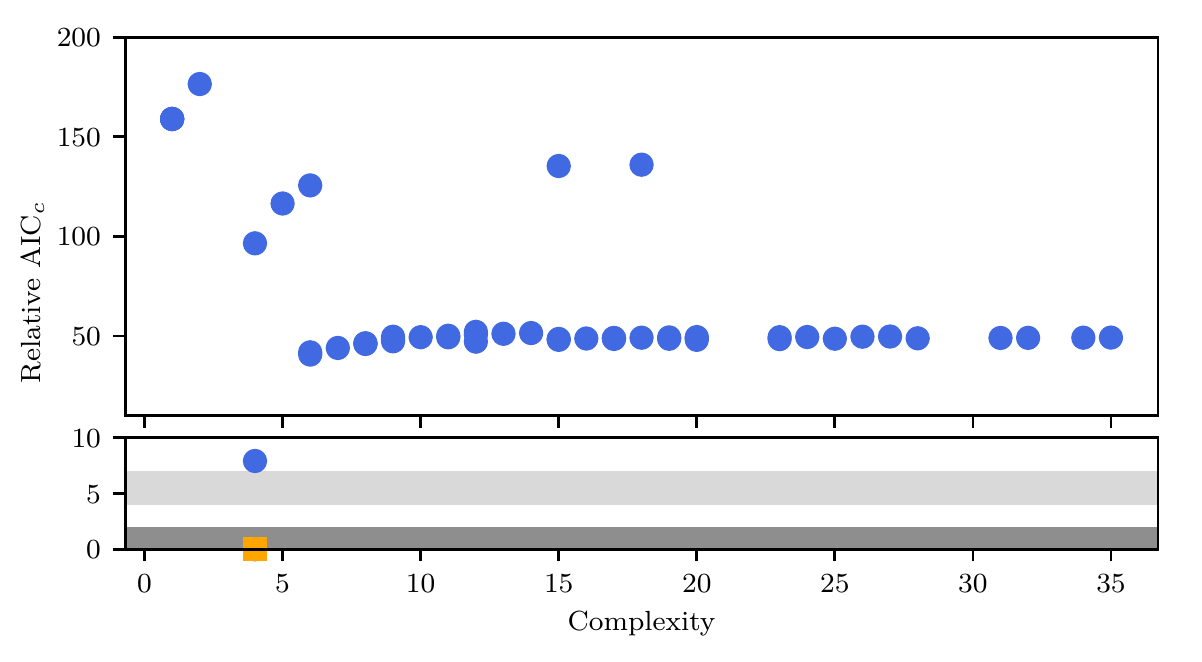}
\caption{Relative AIC\textsubscript{c} criteria for models found by SINDy. The library of polynomial functions used for the identification includes up to 7\textsuperscript{th} degree polynomials in $a_1$ and $a_2$. Magnification in lower panel shows strong (dark gray) and weakly (light gray) supported AIC\textsubscript{c} range. The constrained cubic model ({\color{orange} $\blacksquare$}) identified in \textsection \ref{subsec: dynamical system} lies in the strong support range while its unconstrained counterpart lies just above the weakly supported range.}
\label{fig: model selection}
\end{figure}

Given a library of functions, $\Uptheta(a_1, a_2)$, that includes all polynomials in $a_1$ and $a_2$ up to the 7\textsuperscript{th} degree, figure \ref{fig: model selection} depicts the relative AIC ranking as a function of the model complexity for all of the models identified by SINDy using different sparsity values $\lambda$ to sweep out a Pareto front. Note that, as a starting point, no constraint had been added in the identification step, and the corresponding models are given by the blue dots in figure \ref{fig: model selection}. Under these conditions, the model that optimally balances accuracy and complexity has only 4 terms (blue dot in the lower panel of figure \ref{fig: model selection}) and is given by
\begin{equation}
\begin{aligned}
\frac{\mathrm{d} a_1}{\mathrm{d}t} & = 1.12 a_2 \\
\frac{\mathrm{d} a_2}{\mathrm{d}t} & = -1.116 a_1 + \underbrace{(0.218 - 0.27 a_1^2 - 0.219 a_2^2)}_{2 \sigma(a_1, a_2)} a_2.
\end{aligned}
\label{eq: unconstrained model}
\end{equation}
Although it outperforms all of the other unconstrained models, model \eqref{eq: unconstrained model} suffers a major drawback : the amplitude of the limit cycle it predicts is different from unity (not shown). This misprediction of the amplitude results from the fact that, although the system is supposed to evolve onto the periodic orbit given, due to our normalization, by $a_1^2 + a_2^2 = 1$, the underbraced term in \eqref{eq: unconstrained model} does not vanish. Now knowing the structure of the model, we then add  linear equality constraints in our identification problem \citep{Loiseau2016arxiv} enforcing that the three parameters appearing in the instantaneous growth rate $\sigma(a_1, a_2)$ are equal. Moreover, for the present flow configuration, the leading eigenvalue of the linearized Navier-Stokes operator can be computed by means of an Arnoldi time-stepping algorithm. For the computational domain and mesh considered, the real part of the leading eigenvalue is given by
\begin{equation}
\sigma_{NS} = 0.14.
\notag
\end{equation}
Another linear equality constraint can then be added to the optimization procedure to enforce that the first parameter in \eqref{eq: unconstrained model} is equal to $2 \sigma_{NS}$. The orange square in the lower panel of figure \ref{fig: model selection} corresponds to the model \eqref{eq: lift-based model} presented in \textsection \ref{subsec: dynamical system}. Looking at its AIC ranking, it is clear that the constrained cubic model now outperforms all of the other identified models, even the unconstrained cubic one. This example clearly demonstrates how one can use SINDy \citep{pnas:brunton:2016} and model selection \citep{arxiv:mangan:2016} to identify the optimal structure of the model and then enhance the model performance by adding constraints derived from physical considerations \citep{Loiseau2016arxiv} to the identification procedure.

\section{Identifying a discrete-time dynamical system}
\label{appendix: discrete-time}

\begin{figure}
\centering
\includegraphics[scale=1]{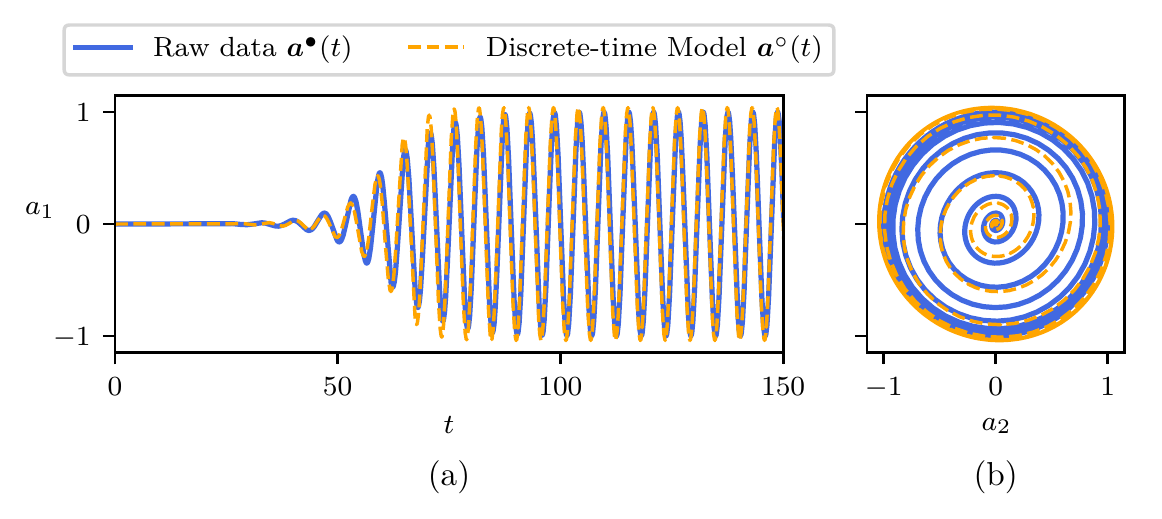}
\caption{(a) Comparison of the evolution as a function of time of the sensor measurement $a_1$ obtained from direct numerical simulation ({\color{blue} \bf -----} $a_1^{\bullet}$) and predicted by the discrete-time sparse model ({\color{orange} \bf -- --} $a_1^{\circ}$). (b) Trajectory of the true and identified systems in the phase plane ($a_1$, $a_2$). In both cases, the initial condition is close to the linearly unstable fixed point.}
\label{fig: appendix -- edmd model}
\end{figure}

In this work, we have used the \emph{Sparse Identification of Nonlinear Dynamics} (SINDy) algorithm, proposed by \cite{pnas:brunton:2016}, in order to identify a reduced-order model of the dynamics in the state ${\bm a}$.
Given time-series of ${\bm a}$ with a sufficiently small sampling period, the original SINDy algorithm allows us to identify the continuous-time dynamical system that governs the dynamics of ${\bm a}$.
A number of alternative system identification techniques could also be used, such as the well-known Volterra series or NARMAX models \citep{Billings2013book,Semeraro2016arxiv,Zhang2012aiaa,Glaz2010aiaa}.
However, these techniques assume a discrete-time representation of the dynamics of the form
\begin{equation}
{\bm a}^{(n+1)} = {\bm f}\left({\bm a}^{(n)} \right).
\notag
\end{equation}
Interestingly, the SINDy algorithm described in \textsection \ref{subsec: SINDy} only requires minor modifications in order to identify such systems. For that purpose, one simply needs to replace the matrix $\dot{\bm A}$ in the optimization problem by a time-shifted copy of ${\bm A}$. Applying this strategy for the two-dimensional cylinder flow at $\Rey=100$ with a time-lag $\tau=0.125$ leads to the identification of the following discrete-time nonlinear dynamical system
\begin{multline}
\begin{bmatrix}
a_1^{(n+1)} \\
a_2^{(n+1)}
\end{bmatrix} = \begin{bmatrix}
						0.994 & 0.139 \\
						-0.122 & 1.026
				      \end{bmatrix} \begin{bmatrix}
												a_1^{(n)} \\
												a_2^{(n)}
											\end{bmatrix} \\+ \begin{bmatrix}
																			0 \\
																			-\left(0.019 a_1^{(n)} + 0.04 a_2^{(n)}\right)\left(a_1^{(n)}\right)^2 - \left(0.017a_1^{(n)} + 0.036a_2^{(n)}\right)\left(a_2^{(n)}\right)^2
																		\end{bmatrix}.
\label{eq: EDMD model}
\end{multline}
The resulting model can be interpreted as a sparse nonlinear vector autoregressive model (VAR) of the first order. As shown in figure \ref{fig: appendix -- edmd model}, the time evolution of $a_1^{\circ}(t)$ predicted by this low-dimensional discrete-time system is in good agreement with $a_1^{\bullet}(t)$ based on a direct numerical simulation, whose initial condition has been chosen in the vicinity of the linearly unstable baseflow.

Based on the continuous-time system \eqref{eq: lift-based model}, it thus appears that SINDy can identify highly accurate low-order models of the two-dimensional cylinder flow.
However, one may argue that direct numerical simulations only provide ideal noise-free and non-corrupted training data, hence questioning the ability of SINDy to identify similar low-dimensional systems from real-world experimental data.
This issue can be addressed by pre-processing the training dataset, such as using a low-pass filter.
Alternatively, \cite{pnas:brunton:2016} used the \emph{Total Variation Regularized Numerical Differentiation} proposed by \cite{Chartrand2011isrnam} in order to evaluate the time-derivative of their noisy data prior to the identification step.
Moreover, the SINDy implementation used by \cite{pnas:brunton:2016}, \cite{Loiseau2016arxiv} and herein relies on an iteratively hard-thresholded least-square algorithm.
As such, one can easily estimate the variance-covariance matrix of the identified model's parameters in order to estimate its robustness even if noisy data are used.
Finally, it has been shown by \cite{arxiv:tran:2016} that SINDy can exactly recover the governing equations even if the training data are highly corrupted, provided the system is sufficiently ergodic or if a sufficient number of different transient trajectories have been included in the training dataset.

%% file: Main.bbl
\begin{thebibliography}{97}
\expandafter\ifx\csname natexlab\endcsname\relax\def\natexlab#1{#1}\fi
\def\au#1{#1} \def\ed#1{#1} \def\yr#1{#1}\def\at#1{#1}\def\jt#1{\textit{#1}}
  \def\bt#1{#1}\def\bvol#1{\textbf{#1}} \def\vol#1{#1} \def\pg#1{#1}
  \def\publ#1{#1}\def\arxiv#1{#1}\def\org#1{#1}\def\st#1{\textit{#1}}

\bibitem[Akaike(1974)]{Akaike1974ieeetac}
{\sc \au{Akaike, H.}} \yr{1974}  \at{A new look at the statistical model
  identification}.  \jt{Automatic Control, IEEE Transactions on}
  \bvol{19}~(6),  \pg{716--723}.

\bibitem[Arbabi \& Mezi{\'c}(2016)]{Arbabi2016arxiv}
{\sc \au{Arbabi, H.} \& \au{Mezi{\'c}, I.}} \yr{2016}  \at{{Ergodic theory,
  Dynamic Mode Decomposition and Computation of Spectral Properties of the
  Koopman operator}}.  \jt{arXiv preprint arXiv:1611.06664} .

\bibitem[Aubry {\em et~al.\/}(1988)Aubry, Holmes, Lumley \&
  Stone]{Aubry1988jfm}
{\sc \au{Aubry, N.}, \au{Holmes, P.}, \au{Lumley, J.~L.} \& \au{Stone, E.}}
  \yr{1988}  \at{The dynamics of coherent structures in the wall region of a
  turbulent boundary layer}.  \jt{J.\ Fluid Mech.}  \bvol{192},  \pg{115--173}.

\bibitem[Babaee \& Sapsis(2016)]{Babaee2016ptrs}
{\sc \au{Babaee, H.} \& \au{Sapsis, T.~P.}} \yr{2016}  \at{A variational
  principle for the description of time-dependent modes associated with
  transient instabilities}.  \jt{Phil.\ Trans.\ Roy.\ S.\ Lond.}
  \bvol{accepted}.

\bibitem[Bagheri(2013)]{jfm:bagheri:2013}
{\sc \au{Bagheri, S.}} \yr{2013}  \at{Koopman-mode decomposition of the
  cylinder wake}.  \jt{J. Fluid Mech}  \bvol{726},  \pg{596--623}.

\bibitem[Balajewicz {\em et~al.\/}(2013)Balajewicz, Dowell \&
  Noack]{Balajewicz2013jfm}
{\sc \au{Balajewicz, M.}, \au{Dowell, E.~H.} \& \au{Noack, B.~R.}} \yr{2013}
  \at{Low-dimensional modelling of high-{R}eynolds-number shear flows
  incorporating constraints from the {N}avier-{S}tokes equation}.  \jt{J.\
  Fluid Mech.}  \bvol{729},  \pg{285--308}.

\bibitem[Barkley \& Henderson(1996)]{jfm:barkley:1996}
{\sc \au{Barkley, D.} \& \au{Henderson, R.~D.}} \yr{1996}
  \at{Three-dimensional {F}loquet stability analysis of the wake of a circular
  cylinder}.  \jt{J. Fluid Mech.}  \bvol{322},  \pg{215--241}.

\bibitem[Barnett \& Hasselmann(1979)]{geophys:barnett:1979}
{\sc \au{Barnett, T.~P.} \& \au{Hasselmann, K.}} \yr{1979}  \at{Techniques of
  linear prediction, with application to oceanic and atmospheric fields in the
  tropical {P}acific}.  \jt{Rev. Geophys.}  \bvol{17},  \pg{949--968}.

\bibitem[Berkooz {\em et~al.\/}(1993)Berkooz, Holmes \&
  Lumley]{berkooz1993proper}
{\sc \au{Berkooz, G.}, \au{Holmes, P.~J.} \& \au{Lumley, J.~L.}} \yr{1993}
  \at{The proper orthogonal decomposition in the analysis of turbulent flows}.
  \jt{Annual review of fluid mechanics}  \bvol{25}~(1),  \pg{539--575}.

\bibitem[Billings(2013)]{Billings2013book}
{\sc \au{Billings, S.~A.}} \yr{2013} {\em Nonlinear system identification:
  {NARMAX} methods in the time, frequency, and spatio-temporal domains\/}.
  \publ{John Wiley \&amp; Sons}.

\bibitem[Bongard \& Lipson(2007)]{Bongard2007pnas}
{\sc \au{Bongard, J.} \& \au{Lipson, H.}} \yr{2007}  \at{Automated reverse
  engineering of nonlinear dynamical systems}.  \jt{Proceedings of the National
  Academy of Sciences}  \bvol{104}~(24),  \pg{9943--9948}.

\bibitem[Bourguet {\em et~al.\/}(2011)Bourguet, Braza \&
  Dervieux]{Bourguet2011jcp}
{\sc \au{Bourguet, R.}, \au{Braza, M.} \& \au{Dervieux, A.}} \yr{2011}
  \at{Reduced-order modeling of transonic flows around an airfoil submitted to
  small deformations}.  \jt{J.\ Comp.\ Phys.}  \bvol{230},  \pg{159--184}.

\bibitem[Brunton {\em et~al.\/}(2017)Brunton, Brunton, Proctor, Kaiser \&
  Kutz]{Brunton2016havok}
{\sc \au{Brunton, S.~L.}, \au{Brunton, B.~W.}, \au{Proctor, J.~L.}, \au{Kaiser,
  E.} \& \au{Kutz, J.~N.}} \yr{2017}  \at{Chaos as an intermittently forced
  linear system}.  \jt{Nature Communications}  \bvol{8}~(19),  \pg{1--9}.

\bibitem[Brunton {\em et~al.\/}(2016{\natexlab{{\em a\/}}})Brunton, Brunton,
  Proctor \& Kutz]{Brunton2016plosone}
{\sc \au{Brunton, S.~L.}, \au{Brunton, B.~W.}, \au{Proctor, J.~L.} \& \au{Kutz,
  J.~N}} \yr{2016{\natexlab{{\em a\/}}}}  \at{Koopman invariant subspaces and
  finite linear representations of nonlinear dynamical systems for control}.
  \jt{PLoS ONE}  \bvol{11}~(2),  \pg{e0150171}.

\bibitem[Brunton {\em et~al.\/}(2014)Brunton, Dawson \& Rowley]{Brunton2014jfs}
{\sc \au{Brunton, S.~L.}, \au{Dawson, S.~TM} \& \au{Rowley, C.~W.}} \yr{2014}
  \at{State-space model identification and feedback control of unsteady
  aerodynamic forces}.  \jt{Journal of Fluids and Structures}  \bvol{50},
  \pg{253--270}.

\bibitem[Brunton \& Noack(2015)]{Brunton2015amr}
{\sc \au{Brunton, S.~L.} \& \au{Noack, B.~R.}} \yr{2015}  \at{Closed-loop
  turbulence control: {P}rogress and challenges}.  \jt{Applied Mechanics
  Reviews}  \bvol{67}~(5),  \pg{050801}.

\bibitem[Brunton {\em et~al.\/}(2016{\natexlab{{\em b\/}}})Brunton, Proctor \&
  Kutz]{pnas:brunton:2016}
{\sc \au{Brunton, S.~L.}, \au{Proctor, J.~L.} \& \au{Kutz, J.~N.}}
  \yr{2016{\natexlab{{\em b\/}}}}  \at{Discovering governing equations from
  data by sparse identification of nonlinear dynamical systems}.
  \jt{Proceedings of the National Academy of Sciences}  \bvol{113}~(15),
  \pg{3932--3937}.

\bibitem[Brunton {\em et~al.\/}(2016{\natexlab{{\em c\/}}})Brunton, Proctor \&
  Kutz]{Brunton2016nolcos}
{\sc \au{Brunton, S.~L.}, \au{Proctor, J.~L.} \& \au{Kutz, J.~N.}}
  \yr{2016{\natexlab{{\em c\/}}}}  \at{{Sparse Identification of Nonlinear
  Dynamics with Control ({SINDYc})}}.  \jt{IFAC NOLCOS}  \bvol{49}~(18),
  \pg{710--715}.

\bibitem[Brunton {\em et~al.\/}(2013)Brunton, Rowley \&
  Williams]{Brunton2013jfm}
{\sc \au{Brunton, S.~L.}, \au{Rowley, C.~W.} \& \au{Williams, D.~R.}} \yr{2013}
   \at{Reduced-order unsteady aerodynamic models at low {R}eynolds numbers}.
  \jt{Journal of Fluid Mechanics}  \bvol{724},  \pg{203--233}.

\bibitem[Cand\`es(2006)]{Candes2006picm}
{\sc \au{Cand\`es, E.~J.}} \yr{2006}  \at{Compressive sensing}.
  \jt{Proceedings of the International Congress of Mathematics} .

\bibitem[Carlberg {\em et~al.\/}(2017)Carlberg, Barone \&
  Antil]{Carlberg2017jcp}
{\sc \au{Carlberg, K.}, \au{Barone, M.} \& \au{Antil, H.}} \yr{2017}
  \at{Galerkin v. least-squares {P}etrov--{G}alerkin projection in nonlinear
  model reduction}.  \jt{Journal of Computational Physics}  \bvol{330},
  \pg{693--734}.

\bibitem[Carlberg {\em et~al.\/}(2015)Carlberg, Tuminaro \&
  Boggs]{Carlberg2015siamjsc}
{\sc \au{Carlberg, K.}, \au{Tuminaro, R.} \& \au{Boggs, P.}} \yr{2015}
  \at{Preserving {L}agrangian structure in nonlinear model reduction with
  application to structural dynamics}.  \jt{SIAM Journal on Scientific
  Computing}  \bvol{37}~(2),  \pg{B153--B184}.

\bibitem[Chartrand(2011)]{Chartrand2011isrnam}
{\sc \au{Chartrand, R.}} \yr{2011}  \at{Numerical differentiation of noisy,
  nonsmooth data}.  \jt{ISRN Applied Mathematics}  \bvol{2011}.

\bibitem[Colebrook(1978)]{ocean:colebrook:1978}
{\sc \au{Colebrook, J.~M.}} \yr{1978}  \at{Continuous plankton records:
  {Z}ooplankton and environment, {N}ortheast {A}tlantic and {N}orth {S}ea}.
  \jt{Oceanol. Acta}  \bvol{1},  \pg{9--23}.

\bibitem[Cordier {\em et~al.\/}(2013)Cordier, Noack, Daviller, Delvile,
  Lehnasch, Tissot, Balajewicz \& Niven]{Cordier2013ef}
{\sc \au{Cordier, L.}, \au{Noack, B.~R.}, \au{Daviller, G.}, \au{Delvile, J.},
  \au{Lehnasch, G.}, \au{Tissot, G.}, \au{Balajewicz, M.} \& \au{Niven, R.K.}}
  \yr{2013}  \at{Control-oriented model identification strategy}.  \jt{Exp.\
  Fluids}  \bvol{54},  \pg{Article 1580}.

\bibitem[Deane {\em et~al.\/}(1991)Deane, Kevrekidis, Karniadakis \&
  Orszag]{Deane1991pfa}
{\sc \au{Deane, A.~E.}, \au{Kevrekidis, I.~G.}, \au{Karniadakis, G.~E.} \&
  \au{Orszag, S.~A.}} \yr{1991}  \at{Low-dimensional models for complex
  geometry flows: Application to grooved channels and circular cylinders}.
  \jt{Phys.\ Fluids A}  \bvol{3},  \pg{2337--2354}.

\bibitem[Donoho(2006)]{Donoho2006ieeetit}
{\sc \au{Donoho, D.~L.}} \yr{2006}  \at{Compressed sensing}.  \jt{IEEE
  Transactions on Information Theory}  \bvol{52}~(4),  \pg{1289--1306}.

\bibitem[Dullerud \& Paganini(2000)]{dp:book}
{\sc \au{Dullerud, Geir.~E.} \& \au{Paganini, F.}} \yr{2000} {\em A course in
  robust control theory: A convex approach\/}.  \publ{Berlin, Heidelberg:
  Springer}.

\bibitem[Duriez {\em et~al.\/}(2016)Duriez, Brunton \&
  Noack]{duriez2016machine}
{\sc \au{Duriez, T.}, \au{Brunton, S.~L.} \& \au{Noack, B.~R.}} \yr{2016}
  \at{{Machine Learning Control--Taming Nonlinear Dynamics and Turbulence}}.
  \jt{Fluid mechanics and its applications (ISSN 0926-5112}  \bvol{116}.

\bibitem[Fabbiane {\em et~al.\/}(2014)Fabbiane, Semeraro, Bagheri \&
  Henningson]{fabbiane2014amr}
{\sc \au{Fabbiane, N.}, \au{Semeraro, O.}, \au{Bagheri, S.} \& \au{Henningson,
  D.~S.}} \yr{2014}  \at{Adaptive and model-based control theory applied to
  convectively unstable flows}.  \jt{Applied Mechanics Reviews}  \bvol{66}~(6),
   \pg{060801}.

\bibitem[Feynman {\em et~al.\/}(2013)Feynman, Leighton \&
  Sands]{Feynman2013book}
{\sc \au{Feynman, R.~P.}, \au{Leighton, R.~B.} \& \au{Sands, M.}} \yr{2013}
  {\em {The Feynman Lectures on Physics}\/}, ,  \vol{vol.~2}.  \publ{Basic
  Books}.

\bibitem[Fischer {\em et~al.\/}(2008)Fischer, Lottes \&
  Kerkemeir]{nek5000_site}
{\sc \au{Fischer, P.F.}, \au{Lottes, J.W.} \& \au{Kerkemeir, S.G.}} \yr{2008}
  {N}ek5000 {W}eb pages. Http://nek5000.mcs.anl.gov.

\bibitem[Galletti {\em et~al.\/}(2004)Galletti, Bruneau, Zannetti \&
  Iollo]{Galletti2004jfm}
{\sc \au{Galletti, G.}, \au{Bruneau, C.~H.}, \au{Zannetti, L.} \& \au{Iollo,
  A.}} \yr{2004}  \at{Low-order modelling of laminar flow regimes past a
  confined square cylinder}.  \jt{J.\ Fluid Mech.}  \bvol{503},  \pg{161--170}.

\bibitem[Ghil {\em et~al.\/}(2002)Ghil, Allen, Dettinger, Ide, Kondrashov {\em
  et~al.\/}]{geophys:ghil:2002}
{\sc \au{Ghil, M.}, \au{Allen, R.~M.}, \au{Dettinger, M.~D.}, \au{Ide, K.},
  \au{Kondrashov, D.} \& \au{others}} \yr{2002}  \at{Advanced spectral methods
  for climatic time series}.  \jt{Rev. Geophys.}  \bvol{40},  \pg{3.1--3.41}.

\bibitem[Glaz {\em et~al.\/}(2010)Glaz, Liu \& Friedmann]{Glaz2010aiaa}
{\sc \au{Glaz, B.}, \au{Liu, L.} \& \au{Friedmann, P.~P.}} \yr{2010}
  \at{Reduced-order nonlinear unsteady aerodynamic modeling using a
  surrogate-based recurrence framework}.  \jt{AIAA journal}  \bvol{48}~(10),
  \pg{2418--2429}.

\bibitem[Graham {\em et~al.\/}(1999)Graham, Peraire \& Tang]{Graham1999ijnme}
{\sc \au{Graham, W.~R.}, \au{Peraire, J.} \& \au{Tang, K.~Y.}} \yr{1999}
  \at{Optimal control of vortex shedding usind low-order models. {P}art {I} ---
  {O}pen-loop model development}.  \jt{Int.\ J.\ Numer.\ Meth. Engrng.}
  \bvol{44},  \pg{945--972}.

\bibitem[Hemati {\em et~al.\/}(2016)Hemati, Dawson \& Rowley]{Hemati2016aiaa}
{\sc \au{Hemati, M.~S.}, \au{Dawson, S.~TM} \& \au{Rowley, C.~W.}} \yr{2016}
  \at{{Parameter-Varying Aerodynamics Models for Aggressive Pitching-Response
  Prediction}}.  \jt{AIAA Journal}  \pg{pp. 1--9}.

\bibitem[Holmes \& Guckenheimer(1983)]{guckenheimer_holmes}
{\sc \au{Holmes, P.~J.} \& \au{Guckenheimer, J.}} \yr{1983} {\em Nonlinear
  oscillations, dynamical systems, and bifurcations of vector fields\/},
  \st{Applied Mathematical Sciences},  \vol{vol.~42}.  \publ{Berlin,
  Heidelberg: Springer-Verlag}.

\bibitem[Holmes {\em et~al.\/}(2012)Holmes, Lumley, Berkooz \&
  Rowley]{HLBR_turb}
{\sc \au{Holmes, P.~J.}, \au{Lumley, J.~L.}, \au{Berkooz, G.} \& \au{Rowley,
  C.~W.}} \yr{2012} {\em Turbulence, coherent structures, dynamical systems and
  symmetry\/}, 2nd edn.  \publ{Cambridge, England: Cambridge University Press}.

\bibitem[Hosseini {\em et~al.\/}(2016)Hosseini, Noack \&
  Martinuzzi]{Hosseini2016jfm}
{\sc \au{Hosseini, Z.}, \au{Noack, B.~R.} \& \au{Martinuzzi, R.~J.}} \yr{2016}
  \at{Modal energy flow analysis of a highly modulated wake behind a
  wall-mounted pyramid}.  \jt{J.\ Fluid Mech.}  \bvol{798},  \pg{774--786}.

\bibitem[Juang \& Pappa(1985)]{jgcd:juang:1985}
{\sc \au{Juang, J.-N.} \& \au{Pappa, R.~S.}} \yr{1985}  \at{An eigensystem
  realization algorithm for modal parameter identification and model
  reduction}.  \jt{Journal of guidance, control, and dynamics}  \bvol{8}~(5),
  \pg{620--627}.

\bibitem[Kaiser {\em et~al.\/}(2014)Kaiser, Noack, Cordier, Spohn, Segond,
  Abel, Daviller, Osth, Krajnovic \& Niven]{Kaiser2014jfm}
{\sc \au{Kaiser, E.}, \au{Noack, B.~R.}, \au{Cordier, L.}, \au{Spohn, A.},
  \au{Segond, M.}, \au{Abel, M.}, \au{Daviller, G.}, \au{Osth, J.},
  \au{Krajnovic, S.} \& \au{Niven, R.~K.}} \yr{2014}  \at{Cluster-based
  reduced-order modelling of a mixing layer}.  \jt{J. Fluid Mech.}  \bvol{754},
   \pg{365--414}.

\bibitem[Kalman(1960)]{Kalman1960jfe}
{\sc \au{Kalman, R.~E.}} \yr{1960}  \at{A new approach to linear filtering and
  prediction problems}.  \jt{Journal of Fluids Engineering}  \bvol{82}~(1),
  \pg{35--45}.

\bibitem[Krizhevsky {\em et~al.\/}(2012)Krizhevsky, Sutskever \&
  Hinton]{Krizhevsky2012nips}
{\sc \au{Krizhevsky, A.}, \au{Sutskever, I.} \& \au{Hinton, G.~E.}} \yr{2012}
  Imagenet classification with deep convolutional neural networks.  \bt{In {\em
  Advances in neural information processing systems\/}},  \pg{pp. 1097--1105}.

\bibitem[Kutz(2017)]{Kutz2017jfm}
{\sc \au{Kutz, J.~N.}} \yr{2017}  \at{Deep learning in fluid dynamics}.
  \jt{Journal of Fluid Mechanics}  \bvol{814},  \pg{1--4}.

\bibitem[Kutz {\em et~al.\/}(2016)Kutz, Brunton, Brunton \&
  Proctor]{Kutz2016book}
{\sc \au{Kutz, J.~N.}, \au{Brunton, S.~L.}, \au{Brunton, B.~W.} \& \au{Proctor,
  J.~L.}} \yr{2016} {\em Dynamic Mode Decomposition: Data-Driven Modeling of
  Complex Systems\/}.  \publ{SIAM}.

\bibitem[Lee {\em et~al.\/}(1997)Lee, Kim, Babcock \& Goodman]{Lee1997pof}
{\sc \au{Lee, C.}, \au{Kim, J.}, \au{Babcock, D.} \& \au{Goodman, R.}}
  \yr{1997}  \at{Application of neural networks to turbulence control for drag
  reduction}.  \jt{Physics of Fluids}  \bvol{9}~(6),  \pg{1740--1747}.

\bibitem[Ling {\em et~al.\/}(2016)Ling, Kurzawski \& Templeton]{Ling2016jfm}
{\sc \au{Ling, J.}, \au{Kurzawski, A.} \& \au{Templeton, J.}} \yr{2016}
  \at{Reynolds averaged turbulence modelling using deep neural networks with
  embedded invariance}.  \jt{Journal of Fluid Mechanics}  \bvol{807},
  \pg{155--166}.

\bibitem[Loiseau \& Brunton(2016)]{Loiseau2016arxiv}
{\sc \au{Loiseau, J.-Ch.} \& \au{Brunton, S.~L.}} \yr{2016}  \at{Constrained
  sparse {Galerkin} regression}.  \jt{arXiv preprint arXiv:1611.03271} .

\bibitem[Mangan {\em et~al.\/}(2016)Mangan, Brunton, Proctor \&
  Kutz]{Mangan2016ieee}
{\sc \au{Mangan, N.~M.}, \au{Brunton, S.~L.}, \au{Proctor, J.~L.} \& \au{Kutz,
  J.~N.}} \yr{2016}  \at{Inferring biological networks by sparse identification
  of nonlinear dynamics}.  \jt{IEEE Transactions on Molecular, Biological, and
  Multi-Scale Communications}  \bvol{2}~(1),  \pg{52--63}.

\bibitem[Mangan {\em et~al.\/}(2017)Mangan, Kutz, Brunton \&
  Proctor]{arxiv:mangan:2016}
{\sc \au{Mangan, N.~M.}, \au{Kutz, J.~N.}, \au{Brunton, S.~L.} \& \au{Proctor,
  J.~L.}} \yr{2017}  \at{Model selection for dynamical systems via sparse
  regression and information criteria}.  \jt{arXiv preprint arXiv:1701.01773} .

\bibitem[Manti{\v{c}}-Lugo {\em et~al.\/}(2014)Manti{\v{c}}-Lugo, Arratia \&
  Gallaire]{prl:mantivc:2014}
{\sc \au{Manti{\v{c}}-Lugo, V.}, \au{Arratia, C.} \& \au{Gallaire, F.}}
  \yr{2014}  \at{Self-consistent mean flow description of the nonlinear
  saturation of the vortex shedding in the cylinder wake}.  \jt{Physical review
  letters}  \bvol{113}~(8),  \pg{084501}.

\bibitem[McConaghy(2011)]{Mcconaghy2011book}
{\sc \au{McConaghy, T.}} \yr{2011}  \at{{Ffx: Fast, scalable, deterministic
  symbolic regression technology}}.  \bt{In {\em Genetic Programming Theory and
  Practice IX\/}},  \pg{pp. 235--260}.  \publ{Springer}.

\bibitem[Mezi{\'c}(2005)]{Mezic2005nd}
{\sc \au{Mezi{\'c}, I.}} \yr{2005}  \at{Spectral properties of dynamical
  systems, model reduction and decompositions}.  \jt{Nonlinear Dynamics}
  \bvol{41}~(1-3),  \pg{309--325}.

\bibitem[Mezi{\'c}(2013)]{Mezic2013arfm}
{\sc \au{Mezi{\'c}, I.}} \yr{2013}  \at{Analysis of fluid flows via spectral
  properties of the {K}oopman operator}.  \jt{Annual Review of Fluid Mechanics}
   \bvol{45},  \pg{357--378}.

\bibitem[Milano \& Koumoutsakos(2002)]{Milano2002jcp}
{\sc \au{Milano, M.} \& \au{Koumoutsakos, P.}} \yr{2002}  \at{Neural network
  modeling for near wall turbulent flow}.  \jt{Journal of Computational
  Physics}  \bvol{182}~(1),  \pg{1--26}.

\bibitem[Nair \& Taira(2015)]{nair2015network}
{\sc \au{Nair, A.~G.} \& \au{Taira, K.}} \yr{2015}  \at{Network-theoretic
  approach to sparsified discrete vortex dynamics}.  \jt{Journal of Fluid
  Mechanics}  \bvol{768},  \pg{549--571}.

\bibitem[Noack(2016)]{Noack2016jfm2}
{\sc \au{Noack, B.~R.}} \yr{2016}  \at{From snapshots to modal expansions --
  bridging low residuals and pure frequencies}.  \jt{J.\ Fluid Mech. -- Focus
  in Fluids}  \bvol{802},  \pg{1--4}.

\bibitem[Noack {\em et~al.\/}(2003)Noack, Afanasiev, Morzynski, Tadmor \&
  Thiele]{jfm:noack:2003}
{\sc \au{Noack, B.~R.}, \au{Afanasiev, K.}, \au{Morzynski, M.}, \au{Tadmor, G.}
  \& \au{Thiele, F.}} \yr{2003}  \at{A hierarchy of low-dimensional models for
  the transient and post-transient cylinder wake}.  \jt{J. Fluid Mech.}
  \bvol{497},  \pg{335--363}.

\bibitem[Noack {\em et~al.\/}(2011)Noack, Morzynski \& Tadmor]{Noack2011book}
{\sc \au{Noack, B.~R.}, \au{Morzynski, M.} \& \au{Tadmor, G.}} \yr{2011} {\em
  Reduced-order modelling for flow control.\/}, ,  \vol{vol. 528}.
  \publ{Springer Science \&amp; Business Media}.

\bibitem[Noack {\em et~al.\/}(2016)Noack, Stankiewicz, Morzy\'nski \&
  Schmid]{Noack2016jfm}
{\sc \au{Noack, B.~R.}, \au{Stankiewicz, W.}, \au{Morzy\'nski, M.} \&
  \au{Schmid, P.~J.}} \yr{2016}  \at{Recursive dynamic mode decomposition of
  transient and post-transient wake flows}.  \jt{J.~Fluid Mech.}  \bvol{809},
  \pg{843--872}.

\bibitem[\"Osth {\em et~al.\/}(2014)\"Osth, Krajnovi\'c, Noack, Barros \&
  Bor\'ee]{Osth2014jfm}
{\sc \au{\"Osth, J.}, \au{Krajnovi\'c, S.}, \au{Noack, B.~R.}, \au{Barros, D.}
  \& \au{Bor\'ee, J.}} \yr{2014}  \at{On the need for a nonlinear subscale
  turbulence term in {POD} models as exemplified for a high {R}eynolds number
  flow over an {A}hmed body}.  \jt{J.\ Fluid Mech.}  \bvol{747},
  \pg{518--544}.

\bibitem[Rediniotis {\em et~al.\/}(2002)Rediniotis, Ko \&
  Kurdila]{Rediniotis2002jfe}
{\sc \au{Rediniotis, O.K.}, \au{Ko, J.} \& \au{Kurdila, A.J.}} \yr{2002}
  \at{Reduced order nonlinear {N}avier-{S}tokes models for synthetic jets}.
  \jt{J.\ Fluids Enrng.}  \bvol{124}~(2),  \pg{433--443}.

\bibitem[Rempfer(2000)]{Rempfer2000tcfd}
{\sc \au{Rempfer, D.}} \yr{2000}  \at{On low-dimensional {G}alerkin models for
  fluid flow}.  \jt{Theoret.\ Comput.\ Fluid Dynamics}  \bvol{14},
  \pg{75--88}.

\bibitem[Rempfer \& Fasel(1994)]{Rempfer1994jfm2}
{\sc \au{Rempfer, D.} \& \au{Fasel, F.~H.}} \yr{1994}  \at{Dynamics of
  three-dimensional coherent structures in a flat-plate boundary-layer}.
  \jt{J.\ Fluid Mech.}  \bvol{275},  \pg{257--283}.

\bibitem[Rowley \& Dawson(2016)]{arfm:rowley:2016}
{\sc \au{Rowley, C.~W.} \& \au{Dawson, S.}} \yr{2016}  \at{Model reduction for
  flow analysis and control}.  \jt{Annual Review of Fluid Mechanics}
  \bvol{49}~(1).

\bibitem[Rowley {\em et~al.\/}(2009)Rowley, Mezi\'c, Bagheri, Schlatter \&
  Henningson]{Rowley2009jfm}
{\sc \au{Rowley, C.~W.}, \au{Mezi\'c, I.}, \au{Bagheri, S.}, \au{Schlatter, P.}
  \& \au{Henningson, D.S.}} \yr{2009}  \at{Spectral analysis of nonlinear
  flows}.  \jt{J. Fluid Mech.}  \bvol{645},  \pg{115--127}.

\bibitem[Rudy {\em et~al.\/}(2017)Rudy, Brunton, Proctor \&
  Kutz]{Rudy2016arxiv}
{\sc \au{Rudy, S.~H.}, \au{Brunton, S.~L.}, \au{Proctor, J.~L.} \& \au{Kutz,
  J.~N.}} \yr{2017}  \at{Data-driven discovery of partial differential
  equations}.  \jt{Science Advances}  \bvol{3}~(e1602614).

\bibitem[Schaeffer(2017)]{Schaeffer2017prsa}
{\sc \au{Schaeffer, H.}} \yr{2017} Learning partial differential equations via
  data discovery and sparse optimization.  \bt{In {\em Proc. R. Soc. A\/}}, ,
  \vol{vol. 473},  \pg{p. 20160446}.

\bibitem[Schlegel \& Noack(2015)]{Schlegel2015jfm}
{\sc \au{Schlegel, M.} \& \au{Noack, B.~R.}} \yr{2015}  \at{On long-term
  boundedness of galerkin models}.  \jt{Journal of Fluid Mechanics}
  \bvol{765},  \pg{325--352}.

\bibitem[Schmid(2010)]{jfm:schmid:2010}
{\sc \au{Schmid, P.~J.}} \yr{2010}  \at{Dynamic mode decomposition of numerical
  and experimental data}.  \jt{Journal of Fluid Mechanics}  \bvol{656},
  \pg{5--28}.

\bibitem[Schmidt \& Lipson(2009)]{Schmidt2009science}
{\sc \au{Schmidt, M.} \& \au{Lipson, H.}} \yr{2009}  \at{Distilling free-form
  natural laws from experimental data}.  \jt{science}  \bvol{324}~(5923),
  \pg{81--85}.

\bibitem[Sch\"olkopf {\em et~al.\/}(1998)Sch\"olkopf, Smola \&
  M\:uller]{Scholkopf1998nc}
{\sc \au{Sch\"olkopf, B.}, \au{Smola, A.} \& \au{M\:uller, K.-R.}} \yr{1998}
  \at{Nonlinear component analysis as a kernel eigenvalue problem}.  \jt{Nerual
  Computation}  \bvol{10},  \pg{1299--1319}.

\bibitem[Schumm {\em et~al.\/}(1994)Schumm, Eberhard \&
  Monkewitz]{jfm:schumm:1994}
{\sc \au{Schumm, M.}, \au{Eberhard, B.} \& \au{Monkewitz, P.~A.}} \yr{1994}
  \at{Self-excited oscillations in the wake of two-dimensional bluff bodies and
  their control}.  \jt{J. Fluid Mech.}  \bvol{271},  \pg{17--53}.

\bibitem[Schwarz {\em et~al.\/}(1978)]{Schwarz1978BIC}
{\sc \au{Schwarz, G.} \& \au{others}} \yr{1978}  \at{Estimating the dimension
  of a model}.  \jt{The annals of statistics}  \bvol{6}~(2),  \pg{461--464}.

\bibitem[Semeraro {\em et~al.\/}(2016)Semeraro, Lusseyran, Pastur \&
  Jordan]{Semeraro2016arxiv}
{\sc \au{Semeraro, O.}, \au{Lusseyran, F.}, \au{Pastur, L.} \& \au{Jordan, P.}}
  \yr{2016}  \at{Qualitative dynamics of wavepackets in turbulent jets}.
  \jt{arXiv preprint arXiv:1608.06750} .

\bibitem[Sengupta {\em et~al.\/}(2015)Sengupta, Haider, Parvathi \&
  Pallavi]{pre:sengupta:2015}
{\sc \au{Sengupta, T.~K.}, \au{Haider, S.~I.}, \au{Parvathi, M.~K.} \&
  \au{Pallavi, G.}} \yr{2015}  \at{Enstrophy-based proper orthogonal
  decomposition for reduced-order modeling of flow past a cylinder}.
  \jt{Physical Review E}  \bvol{91}~(4),  \pg{043303}.

\bibitem[Sipp \& Schmid(2016)]{Sipp2016amr}
{\sc \au{Sipp, D.} \& \au{Schmid, P.~J.}} \yr{2016}  \at{Linear closed-loop
  control of fluid instabilities and noise-induced perturbations: {A} review of
  approaches and tools}.  \jt{Applied Mechanics Reviews}  \bvol{68}~(2),
  \pg{020801}.

\bibitem[Sirovich(1987)]{qam:sirovich:1987}
{\sc \au{Sirovich, L.}} \yr{1987}  \at{Turbulence and the dynamics of coherent
  structures. {P}art {I}: {C}oherent structures}.  \jt{Quarterly of Applied
  Mathematics}  \bvol{45}~(3),  \pg{561--571}.

\bibitem[Skogestad \& Postlethwaite(2005)]{sp:book}
{\sc \au{Skogestad, S.} \& \au{Postlethwaite, I.}} \yr{2005} {\em Multivariable
  feedback control: analysis and design\/}, 2nd edn.  \publ{Hoboken, New
  Jersey: John Wiley \& Sons, Inc.}

\bibitem[Tadmor {\em et~al.\/}(2011)Tadmor, Lehmann, Noack, Cordier, Delville,
  Bonnet \& Morzy\'nski]{Tadmor2011ptrsa}
{\sc \au{Tadmor, G.}, \au{Lehmann, O.}, \au{Noack, B.~R.}, \au{Cordier, L.},
  \au{Delville, J.}, \au{Bonnet, J.-P.} \& \au{Morzy\'nski, M.}} \yr{2011}
  \at{Reduced order models for closed-loop wake control}.  \jt{Philosophical
  Transactions of the Royal Society A}  \bvol{369}~(1940),  \pg{1513--1524}.

\bibitem[Tadmor {\em et~al.\/}(2010)Tadmor, Lehmann, Noack \&
  Morzy{\'n}ski]{pof:tadmor:2010}
{\sc \au{Tadmor, G.}, \au{Lehmann, O.}, \au{Noack, B.~R.} \& \au{Morzy{\'n}ski,
  M.}} \yr{2010}  \at{Mean field representation of the natural and actuated
  cylinder wake}.  \jt{Physics of Fluids (1994-present)}  \bvol{22}~(3),
  \pg{034102}.

\bibitem[Takens(1981)]{takens:1981}
{\sc \au{Takens, F.}} \yr{1981}  \at{Detecting strange attractors in
  turbulence}.  \bt{In {\em Dynamical systems and turbulence, Warwick 1980\/}},
   \pg{pp. 366--381}.  \publ{Springer}.

\bibitem[Tibshirani(1996)]{Tibshirani1996lasso}
{\sc \au{Tibshirani, R.}} \yr{1996}  \at{Regression shrinkage and selection via
  the lasso}.  \jt{Journal of the Royal Statistical Society. Series B
  (Methodological)}  \pg{pp. 267--288}.

\bibitem[{Tran} \& {Ward}(2016)]{arxiv:tran:2016}
{\sc \au{{Tran}, G.} \& \au{{Ward}, R.}} \yr{2016}  \at{{Exact Recovery of
  Chaotic Systems from Highly Corrupted Data}}.  \jt{ArXiv e-prints} ,
  \arxiv{arXiv: 1607.01067}.

\bibitem[Tu {\em et~al.\/}(2014)Tu, Rowley, Luchtenburg, Brunton \&
  Kutz]{Tu2014jcd}
{\sc \au{Tu, J.~H.}, \au{Rowley, C.~W.}, \au{Luchtenburg, D.~M.}, \au{Brunton,
  S.~L.} \& \au{Kutz, J.~N.}} \yr{2014}  \at{On dynamic mode decomposition:
  theory and applications}.  \jt{Journal of Computational Dynamics}
  \bvol{1}~(2),  \pg{391--421}.

\bibitem[Ukeiley {\em et~al.\/}(2001)Ukeiley, Cordier, Manceau, Delville,
  Bonnet \& Glauser]{Ukeiley2001jfm}
{\sc \au{Ukeiley, L.}, \au{Cordier, L.}, \au{Manceau, R.}, \au{Delville, J.},
  \au{Bonnet, J.~P.} \& \au{Glauser, M.}} \yr{2001}  \at{Examination of
  large-scale structures in a turbulent plane mixing layer. {P}art 2.
  {D}ynamical systems model}.  \jt{J.\ Fluid Mech.}  \bvol{441},  \pg{61--108}.

\bibitem[Wang {\em et~al.\/}(2011)Wang, Yang, Lai, Kovanis \&
  Grebogi]{Wang2011prl}
{\sc \au{Wang, W.~X.}, \au{Yang, R.}, \au{Lai, Y.~C.}, \au{Kovanis, V.} \&
  \au{Grebogi, C.}} \yr{2011}  \at{Predicting catastrophes in nonlinear
  dynamical systems by compressive sensing}.  \jt{Physical Review Letters}
  \bvol{106},  \pg{154101--1--154101--4}.

\bibitem[Weare \& Nasstrom(1982)]{weathervec:weare:1982}
{\sc \au{Weare, B.~C.} \& \au{Nasstrom, J.~N.}} \yr{1982}  \at{Examples of
  extended empirical orthogonal function analyses}.  \jt{Mon. Weather Rev.}
  \bvol{110},  \pg{784--812}.

\bibitem[Wei \& Rowley(2009)]{Wei2009jfm}
{\sc \au{Wei, M.} \& \au{Rowley, C.~W.}} \yr{2009}  \at{Low-dimensional models
  of a temporally evolving free shear layer}.  \jt{J.\ Fluid Mech}  \bvol{618},
   \pg{113--134}.

\bibitem[Welch \& Bishop(1995)]{Welch1995book}
{\sc \au{Welch, G.} \& \au{Bishop, G.}} \yr{1995} {An introduction to the
  Kalman filter}.

\bibitem[Wiener(1948)]{Wiener1948book}
{\sc \au{Wiener, N.}} \yr{1948} {\em Cybernetics or Control and Communication
  in the Animal and the Machine\/}, 1st edn.  \publ{Boston: MIT Press}.

\bibitem[Williams {\em et~al.\/}(2015)Williams, Kevrekidis \&
  Rowley]{Williams2015jnls}
{\sc \au{Williams, M.~O.}, \au{Kevrekidis, I.~G.} \& \au{Rowley, C.~W.}}
  \yr{2015}  \at{A data-driven approximation of the {K}oopman operator:
  extending dynamic mode decomposition}.  \jt{Journal of Nonlinear Science} .

\bibitem[Zebib(1987)]{jem:zebib:1987}
{\sc \au{Zebib, A.}} \yr{1987}  \at{Stability of viscous flow past a circular
  cylinder}.  \jt{Journal of Engineering Mathematics}  \bvol{21}~(2),
  \pg{155--165}.

\bibitem[Zhang {\em et~al.\/}(1995)Zhang, Fey, Noack, K{\"o}nig \&
  Eckelmann]{pof:zhang:1995}
{\sc \au{Zhang, H.-Q.}, \au{Fey, U.}, \au{Noack, B.~R.}, \au{K{\"o}nig, M.} \&
  \au{Eckelmann, H.}} \yr{1995}  \at{On the transition of the cylinder wake}.
  \jt{Physics of Fluids (1994-present)}  \bvol{7}~(4),  \pg{779--794}.

\bibitem[Zhang {\em et~al.\/}(2012)Zhang, Wang, Ye \& Quan]{Zhang2012aiaa}
{\sc \au{Zhang, W.}, \au{Wang, B.}, \au{Ye, Z.} \& \au{Quan, J.}} \yr{2012}
  \at{Efficient method for limit cycle flutter analysis based on nonlinear
  aerodynamic reduced-order models}.  \jt{AIAA journal}  \bvol{50}~(5),
  \pg{1019--1028}.

\bibitem[Zhang \& Duraisamy(2015)]{Zhang2015aiaa}
{\sc \au{Zhang, Z.~J.} \& \au{Duraisamy, K.}} \yr{2015} Machine learning
  methods for data-driven turbulence modeling.  \bt{In {\em 22nd AIAA
  Computational Fluid Dynamics Conference\/}},  \pg{p. 2460}.

\end{thebibliography}
